\documentclass{article}

\usepackage{tikz}
\usetikzlibrary{positioning}
\usetikzlibrary{calc,trees,positioning,arrows,chains,shapes.geometric,%
    decorations.pathreplacing,decorations.pathmorphing,shapes,%
    matrix,shapes.symbols,automata}

\usetikzlibrary{fit}
\usetikzlibrary{shapes}
\usetikzlibrary{decorations.markings}

\tikzstyle{randomVariable}=[circle,fill=white,draw=black,text=black,minimum size=0.8cm]
\tikzstyle{param}=[draw=none,fill=none,text=black,minimum size=0.2cm]

\usepackage[american]{babel}
\usepackage{epsfig}
\usepackage{color}
\usepackage{amsmath, amssymb, amsfonts, amstext, latexsym}
\usepackage{yhmath}
\usepackage{stmaryrd}
\allowdisplaybreaks
\newtheorem{theorem}{Theorem}
\newtheorem{lemma}{Lemma}
\newtheorem{remark}{Remark}
\newtheorem{proof}{Proof}
\newtheorem{proposition}{Proposition}

\newtheorem{definition}{Definition}
\newtheorem{assumption}{Assumption}
\newcommand{\V}{\mathcal{V}}
\newcommand{\Ecal}{\mathcal{E}}

\newcommand{\E}{\mathbb{E}}
\newcommand{\N}{\mathbb{N}}

\newcommand{\I}{\mathcal{I}}
\newcommand{\Iset}{\mathbb{I}}

\usepackage{hyperref}
\usepackage{url}

\AtBeginDocument{%
  \providecommand\BibTeX{{%
    \normalfont B\kern-0.5em{\scshape i\kern-0.25em b}\kern-0.8em\TeX}}}

\begin{document}

\title{Flexibility can hurt dynamic matching system performance}


\author{Arnaud Cadas\thanks{INRIA, Paris, France. DI ENS, CNRS, PSL Research University, Paris, France.} 
\and Josu Doncel\thanks{University of the Basque Country, UPV-EHU, Leioa, Spain.} \and Jean-Michel Fourneau\thanks{DAVID, UVSQ, Université Paris-Saclay, Versailles, France.} 
\and Ana Bu\v{s}i\'c\footnotemark[1]}

\maketitle

\begin{abstract}
We study the performance of general dynamic matching models. This model is defined by a connected graph, where nodes represent the class of items and the edges the compatibilities between items. Items of different classes arrive one by one to the system according to a given probability distribution. Upon arrival, an item is matched with a compatible item according to the First Come First Served discipline and leave the system immediately, whereas it is enqueued with other items of the same class, if any. We show that such a model may exhibit a non intuitive behavior: increasing the services ability by adding new edges in the matching graph may lead to a larger average population. This is similar to a Braess paradox. We first consider a quasicomplete graph with four nodes and we provide values of the probability distribution of the arrivals such that when we add an edge the mean number of items is larger. Then, we consider an arbitrary matching graph and we show sufficient conditions for the existence or non-existence of this paradox. We conclude that the analog to the Braess paradox in matching models is given when specific independent sets are in saturation, i.e., the system is close to the stability condition.
\end{abstract}


\newpage
\section{Introduction}

The Braess paradox is possibly one 
of the most important results in network modelling of the last decades.
It states that, when the agents that participate in a traffic network can
take self-interested decisions, the running times of the agents can increase if we add a new road. 
The idea behind this phenomenon is that the extension of the network might cause a
redistribution of the traffic that increases the congestion and, as a result, the delay of agents.
More precisely, the Braess paradox shows that the travel time in the Nash equilibrium (the set of strategies such as no agent has incentive to deviate unilaterally) 
can increase if we add a shortcut in the network. This result reflects that 
the selfish behavior of agents in a network might lead to 
a situation whose performance is not the optimal one or, in other words, 
that the Price of Anarchy (the ratio between the performance of the system in the Nash equilibrium
over the optimal performance) is larger than one.

A matching model is defined by a set of items classes and a set of compatibilities
among classes of items that determine the classes of items that can be matched with each other. 
Items of different classes arrive one by one to the system according to a given probability
distribution. Upon arrival, an item is queued if there are not 
compatible items present in the system. However, if there are compatible items, it
is matched with one of its compatible items and then both items leave the system immediately. 
A matching policy determines how compatible items are matched.
Some examples of matching policies are First Come First Served (FCFS), where the incoming item is matched 
with the oldest compatible item, or Match the Longest, where  item is 
matched with the compatible class with the largest number of items present.

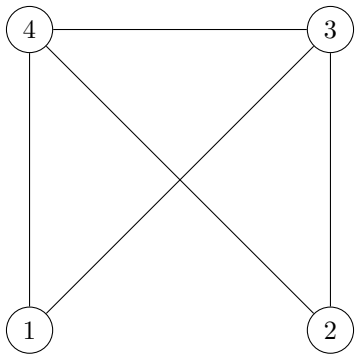
\begin{figure}[htbp]
     \centering
    \begin{tikzpicture}[]
     \node[style={circle,draw}] at (0,0) (1) {$1$};
     \node[style={circle,draw}] at (4,0) (2) {$2$};
     \node[style={circle,draw}] at (4,4) (3) {$3$};
     \node[style={circle,draw}] at (0,4) (4) {$4$};
     \draw (4) -- (2);
     \draw (1) -- (3);
     \draw (2) -- (3);
     \draw (3) -- (4);
     \draw (4) -- (1);
    \end{tikzpicture}  
   \caption{Example of a compatibility graph for a matching model.}
   \label{fig:4nodes-delta-intro}
  \end{figure}

Let us present an example of a matching model. We consider the compatibility graph of
Fig.~\ref{fig:4nodes-delta-intro}. As it can be observed, there are four classes of items. Items
of class 3 and 4 can be matched with all the classes, whereas items of class 1 and class 2 can only be matched with items of class 3 and class 4.

\begin{figure*}[t!]
\centering
\includegraphics[width=1\columnwidth]{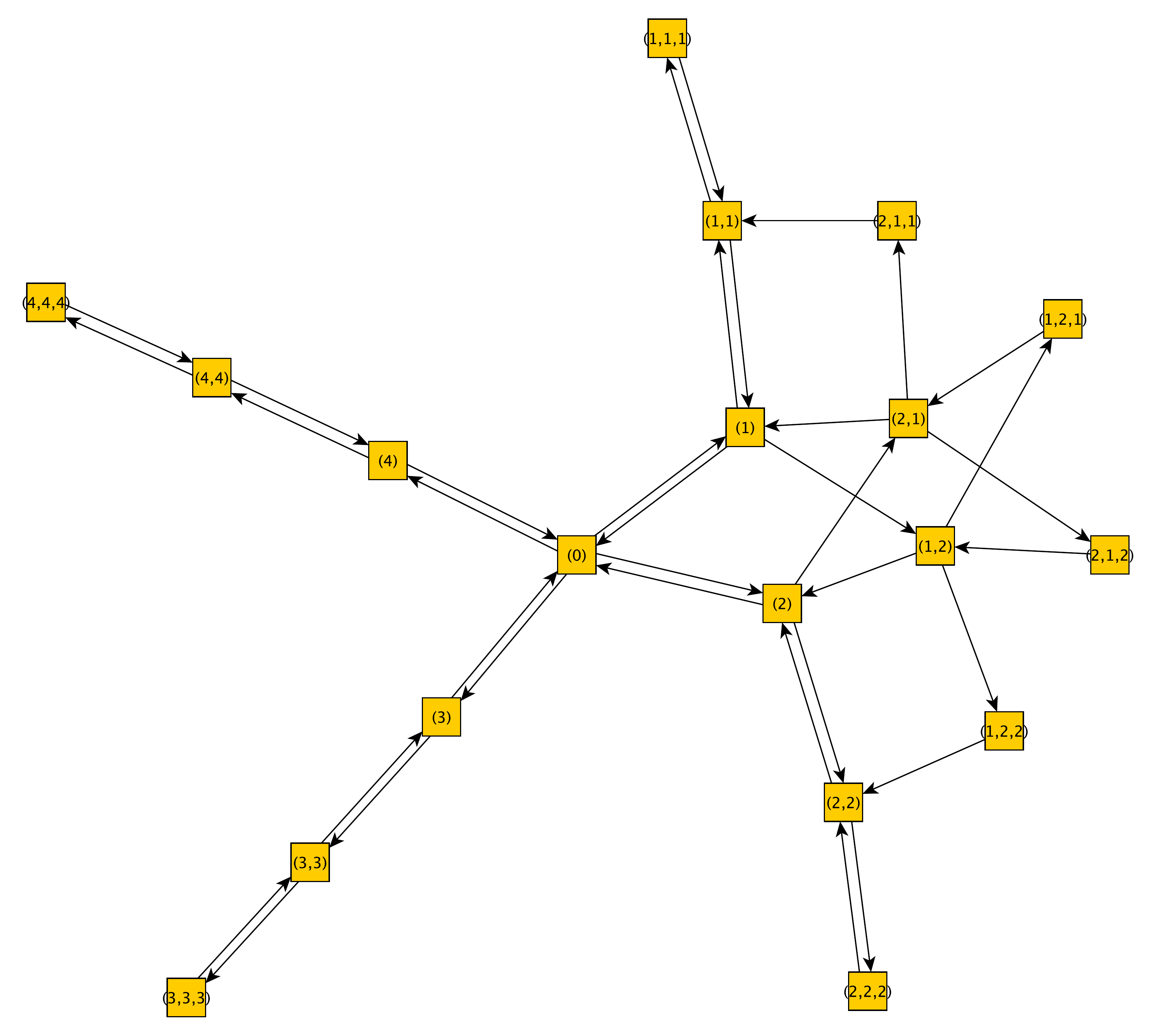}
\caption{The Markov Chain (truncated to words of size 3) derived from the matching model of
Fig.~\ref{fig:4nodes-delta-intro}.}
\label{fig:g4l}
\end{figure*}

We would like to remark that, when the matching policy is FCFS and the arrivals are $i.i.d$, the dynamics of this 
matching model is described by the Discrete Time Markov Chain of Fig.~\ref{fig:g4l}. Note
that we represent only the states with three of less elements in this illustration. 
In this Markov Chain, the state $(i,j,k)$ represents that there are three items in the system
and item $i$ arrived first, then item $j$ and finally item $k$. We now provide a trajectory of
the random variable of the items remaining after the matchings, when at the initial time the system is empty.


\begin{itemize}
\item \textbf{Arrival of an item of class 1}. When the system is empty and there is an arrival 
of an item of class 1 it is buffered. The state of the system is (1).
\item \textbf{Arrival of an item of class 2}. When the state of the system is (1) and an item 
of class 2 arrives, the incoming item is buffered since it can not be matched with the 
item of class 1. The state is (1,2).
\item \textbf{Arrival of an item of class 1}. We now consider that an item of class 1
arrives when the state is (1,2). Hence, 
the incoming item is also buffered since it can not be matched with the classes of items 
that are present. Therefore, the state of the system is (1,2,1).
\item \textbf{Arrival of an item of class 3}. We consider that an item of class 3 
arrives to the system when the state is (1,2,1). 
The incoming item can be matched with both classes of items. Since
the matching policy is FCFS, the incoming item is matched with an item of class 1 
and both items leave. As a result, the state of the system is (2,1).
\item \textbf{Arrival of an item of class 4}. We consider that an item of class 4 
arrives to the system when the state is (2,1). 
Since the item of class 2 arrived first, it is matched with the incoming item and they
depart. Therefore, the state of the system is (1).
\end{itemize}

We aim to investigate the existence of a Braess paradox in matching models, i.e,  
we study the influence on the mean number of items when we add an edge
in a matching model. 
In \cite{Long}, the authors show that the optimal matching policy (i.e., that minimizes the mean number of items
in the system) is a stationary policy where each matching is the solution of the Bellman fixed point equation, i.e each matching is the minimizer of the Bellman operator among all possible matchings. Adding an edge to the compatibility graph leads to a bigger set of possible matchings which means that the minimum can only be smaller. Therefore, it is clear that, when we add an edge to the compatibility graph, the performance of the system
will always improve under the optimal policy.
However, this is less obvious for other matching policies. 
This is, in fact, the problem we address in this article, that is, we study the conditions under which
the performance of a matching model can be worsen when we add an edge to the compatibility
graph. 

We consider that items are matched according to the FCFS policy. This assumption allows us to
use the result of \cite{MBM20} that shows that the steady-state
distribution of items  admits a product form solution when the matching discipline
is FCFS. 

The main contributions of our paper are the following:

\begin{itemize}
\item We first consider the matching model of Fig.~\ref{fig:4nodes-delta-intro} and we
provide necessary and sufficient conditions on the arrivals for the existence of a Braess paradox.
The intuition behind the existence of a Braess paradox is the following: when node 3 is close to saturation, i.e., the difference
between the arrivals to node 3 and the sum of arrivals of the rest of the nodes is small, the 
number of items in that node grows large. Thus, when we add the edge (1,2), we allow items
of class 1 and of class 2 to be matched and, therefore, they are not always matched with items
of class 3, which causes that the number of items of class 3 increases.
\item We analyze an arbitrary matching model and we study the 
existence of a Braess paradox for this model. First, we show that the mean number of
items in the system can be written as a finite sum over all independent sets. Then, we give sufficient conditions
for the existence or the non-existence of a Braess paradox under the assumption of saturation for one independent set. We prove that a Braess paradox exists when an independent set, that does not 
contain any of the nodes of the edge we add but has them as neighbors, is in saturation. We also prove that a Braess paradox does not exist when an independent set that 
is in saturation contains one of the nodes of the edge we add.  
\item We also study the influence on the mean number of items of adding new item classes 
in the matching model. We show that if we add a class with the same neighborhood of 
a class that is present in the system, the arrivals can be set in a way that the mean number
of items does not change.  This result allows us conclude that we can construct a huge number
of matching models where a Braess paradox exists, from a matching model such that the mean
number of items increases when we add an edge.
\end{itemize}

The rest of the article is organized as follows: we present some related work about matching models, FCFS policy and Braess paradox in Section~\ref{sec:related_work}. In Section~\ref{sec:model}, we describe more formally the model and introduce useful notations. A first example of the Braess paradox in matching models is shown in Section~\ref{sec:quasi_complete} using a quasicomplete compatibility graph with four nodes. Then, sufficient conditions for the existence or non-existence of the Braess paradox in arbitrary graphs are given in Section~\ref{sec:arbitrary_graph}. Section~\ref{sec:extensions} is devoted to present the extensions of a compatibility graph such that a Braess paradox still holds in the new matching model. Finally, we discuss future work in Section~\ref{sec:conclusion}.

\section{Related Work}\label{sec:related_work}



The study of dynamic matching models has recently gained a lot of interest of researchers in
different areas due to its applications in organ donation \cite{KidneySite,Kidney}, ridesharing \cite{BKQ}, power grid \cite{ZDC} 
or pattern recognition \cite{SR}. In this context,
there is a wide range of papers that investigate bipartite matching models where the classes of items can be divided into two groups, the compatibility graph is bipartite and items arrive as a couple of one of each group.
In \cite{Stability}, the authors provide necessary conditions for the stability of these matching 
models. The FCFS infinite bipartite matching model was introduced in \cite{FCFSModel}
and the existence of the product form of the stationary distribution has been shown in \cite{AW,ExactFCFS}, whereas in \cite{ADAN2018253} the authors show how it is closely related
to the model of \cite{gardner2016queueing} with redundant requests. 
When holding costs in the buffers are considered, the optimal policy of 
the bipartite matching model has been studied by \cite{BoundedRegret} in an asymptotic regime
and by \cite{Long} in a non-asymptotic one. In \cite{GW}, the authors present the
imbalance process and derived a lower bound on the holding costs.

In our paper, we consider matching models with non-bipartite compatibility graphs where items arrive one by one. This kind of models has been also considered in the literature. The authors in \cite{Mairesse_Stability} study the stability of non-bipartite matching models. In \cite{MBM20} the authors show that the existence of the product form
of the stationary distribution when the items are matched according to the FCFS policy.

The existence of a
Braess paradox has been explored in several contexts related to queueing networks 
(see for instance \cite{BKT97,calvert_solomon_ziedins_1997,cohen1997congestion,cohen1990paradox,kameda2002harmful}).
However, to the best of our knowledge, it has not been studied so far in matching models.

\section{Model Description}\label{sec:model}

We consider the following matching model in discrete time. 
In each time slot $n\in\N^\ast$, one item arrives to the system with probability $1-\alpha_0$ (and nothing happen with probability $\alpha_0$). We assume that this item belongs to one class within the set of classes denoted by $\mathcal V=\{1,\dots,n\}$. Its class is chosen at random given a probability distribution $\alpha=(\alpha_1,\cdots,\alpha_n)$ over $\V$.
Then, the item is matched according to a policy using the non-bipartite compatibility graph $\mathcal{G}=(\mathcal V,\xi)$ (we consider that every compatibility graph is non-bipartite and we will not precise it for the remainder of the paper), 
where $\mathcal{V}$ is the set of nodes (each node represents one class) and $\xi$ is the set of allowed matching pairs. Two items of classes $i$ and $j$ are compatible if and only if $(i,j)\in\xi$. If the incoming item is compatible with at least one other item already in the system, we use the FCFS policy which matches the former with the oldest compatible item. Otherwise, the incoming item is placed at the end of the queue related to its class.

Let $\V^\ast$ be the set of finite words over the alphabet $\V$ and $\mathbb{W}=\{w=w_1 \cdots w_q\in\V^\ast : \forall (i,j)\in\llbracket 1,q \rrbracket^2,\; i\neq j,\; (w_i,w_j)\notin \xi\}$ be the subset of words such that there is no pair of letters that are compatible.
A state of the system just after having done the matching and before the next incoming item can be represented by a word $w=w_1 \cdots w_q \in\mathbb{W}$. Each letter $w_i\in\V$ represent the class of an item remaining in the system and the order of the letters represent the order of arrival of these items.

The whole process can be modeled as a discrete time Markov chain $W=(W_n)_{n\in\N}$ with values in $\mathbb{W}$, an initial state $w^0$ and with the following transitions: assume $W_n = w=w_1\cdots w_q\in\mathbb{W}$, if an item of class $i\in\V$ arrives (with probability $\alpha_i$), then
\[
W_{n+1}=\left\{\begin{array}{lr}
w i  & \text{if }\forall l\in\{1,\cdots,q\}, w_l\notin\Ecal(i) \\
w_{-i} & \text{Otherwise}
\end{array}\right.
\]
with $w_{-i}$ being defined as the word $w$ where we removed the first appearance of any letter that belongs to $\Ecal(i)$ and $\Ecal(i)=\{j\in\V : (i,j)\in\xi\}$ (i.e $\Ecal(i)$ is the set of all the neighbors of the node $i$).

Let $\I$ be an independent set of $\mathcal G$ (i.e a non-empty subset of $\V$ such that any pair of nodes are not linked together) and $\Iset$ be the set of independent sets of $\mathcal G$. We define $|\alpha_V|=\sum_{i\in V}\alpha_i$ for any subset $V\subseteq \V$.
We assume that $\alpha$ is chosen such that it satisfies the necessary
and sufficient conditions for stabilizability of the model (i.e there exist a matching policy under which the Markov chain is positive recurrent): \texttt{Ncond} given in \cite{Mairesse_Stability}, i.e
\begin{equation} \label{eq:stability}
|\alpha_{\I}|<|\alpha_{\Ecal(\I)}|\quad \forall \I\in\Iset
\end{equation}
where $\Ecal(V)=\bigcup_{i\in V}\Ecal(i)$ for any subset $V\subseteq\V$.

In \cite{MBM20} the authors prove that under \texttt{Ncond}, the FCFS policy is stable (i.e $W$ is positive recurrent) and give the stationary distribution of $W$. We recall its value in the following proposition because it will be the foundation of all the results throughout this paper.

\begin{proposition}[Theorem 1,\cite{MBM20}]\label{prop:stationary_dist}
Assume that $\alpha$ satisfy \eqref{eq:stability}. Then, the stationary distribution of $W$, noted $\pi$, is equal to
\[\pi(w)=\pi_0\prod_{i=1}^q \frac{\alpha_{w_i}}{|\alpha_{\Ecal(\{w_1,\cdots, w_i\})}|}\quad ,\text{ for any }w=w_1\cdots w_q\in\mathbb{W},\]
where $\pi_0$ is the normalization constant.
\end{proposition}

Let $w\in\mathbb{W}$, we denote by $|w|_i$ the number of times the letter $i$ appears in the word $w$ for any $i\in\V$, i.e the number of items of class $i$ remaining in the system. We define $Q_n =\sum_{i\in\V}|W_n|_i$ as the total number of items remaining in the system at time $n$. We denote by $\mathbb E[Q]$ the mean total number of items present in the system under the stationary distribution $\pi$. 

We also consider another compatibility graph $\overline{\mathcal G}=(\V,\overline{\xi})$ where we added the edge $(i^\ast,j^\ast)$, i.e $\overline{\xi}=\xi\cup\{(i^\ast,j^\ast)\}$. We denote by $\overline{W}$ the Markov chain defined on $\overline{\mathcal{G}}$ and by $\mathbb E[\overline Q]$ the mean total number of items present in the system with the added edge.

We say that there exists a Braess paradox if
$$
\mathbb E[\overline Q]>\mathbb E[Q],
$$
that is, when the mean number of items increases if we add an edge to the matching model.

\section{Quasicomplete Graph with Four Nodes}\label{sec:quasi_complete}

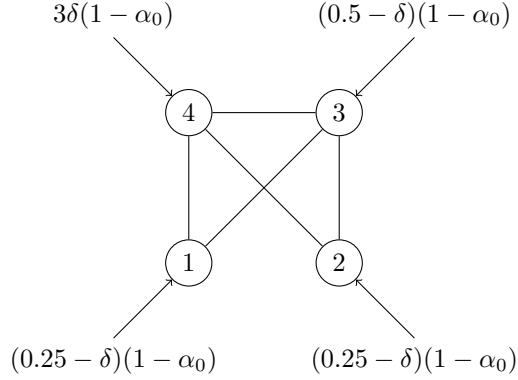
\begin{figure}[htbp]
     \centering
    \begin{tikzpicture}[]
     \node[style={circle,draw}] at (0,0) (1) {$1$};
     \node[style={circle,draw}] at (2,0) (2) {$2$};
     \node[style={circle,draw}] at (2,2) (3) {$3$};
     \node[style={circle,draw}] at (0,2) (4) {$4$};
     \draw[<-] (1) -- (-1,-1)  node[below] {$(0.25-\delta)(1-\alpha_0)$} ;
    \draw[<-] (2) -- (3,-1)  node[below] {$(0.25-\delta)(1-\alpha_0)$} ;
    \draw[<-] (3) -- (3,3)  node[above] {$(0.5-\delta)(1-\alpha_0)$} ;
    \draw[<-] (4) -- (-1,3)  node[above] {$3\delta(1-\alpha_0)$} ;
     \draw (4) -- (2);
     \draw (1) -- (3);
     \draw (2) -- (3);
     \draw (3) -- (4);
     \draw (4) -- (1);
    \end{tikzpicture}  
   \caption{A matching graph that consists of a quasi-complete graph with $4$ nodes.}
   \label{fig:4nodes-delta}
  \end{figure}

We study the existence of a Braess paradox in a matching model whose compatibility 
graph is a quasicomplete graph formed by four nodes (see Fig.~\ref{fig:4nodes-delta}). 
In this section, we provide necessary and sufficient conditions on the arrivals for the existence of 
a Braess paradox in this model, that is, the conditions on the arrivals under which the mean
number of items increases if we add the edge (1,2). 

We note that the matching model of Fig.~\ref{fig:4nodes-delta} is the same than that of
Fig.~\ref{fig:4nodes-delta-intro} with the following probability distribution of the arrivals: 
$\alpha_1=\alpha_2=0.25-\delta$, $\alpha_3=0.5-\delta$
and $\alpha_4=3\delta.$ Throughout this section, we assume that 
$\delta\in (0,\tfrac{1}{6})$ to ensure that \eqref{eq:stability} is satisfied.

In the following
result, we provide 
an expression of the mean number of items for this model.

\begin{lemma}
For the matching model of Fig.~\ref{fig:4nodes-delta}, the mean number of items is
$$
\frac{\frac{(0.5-2\delta)(0.5+2\delta)}{(4\delta)^2}+\frac{(0.5-\delta)(0.5+\delta)}{(2\delta)^2}+\frac{3\delta(1-3\delta)}{(1-6\delta)^2}}
{1+\frac{0.5-2\delta}{4\delta}+\frac{0.5-\delta}{2\delta}+\frac{3\delta}{1-6\delta}}.
$$
\label{lem:en-4nodes-delta-quasi}
\end{lemma}

\begin{proof}
Since node $1$ and node $2$ have the same neighbors, we 
use lumpability to aggregate these nodes. We denote by $A$ the aggregated node and 
the arrival rate to the node A by $\alpha_A=\alpha_1+\alpha_2$. Therefore,
the derived graph is a complete graph with $3$ nodes, $A,3,4$.

Then, we use Proposition~\ref{prop:finite_sums} where $\Iset=\{\{A\}, \{3\}, \{4\}\}$, $\alpha_A=0.5-2\delta$, $\alpha_3=0.5-\delta$, $\alpha_4=3\delta$, $|\alpha_{\Ecal(\{A\})}|=0.5+2\delta$, $|\alpha_{\Ecal(\{3\})}|=0.5+\delta$
and $|\alpha_{\Ecal(\{4\})}|=1-3\delta$. 
\end{proof}

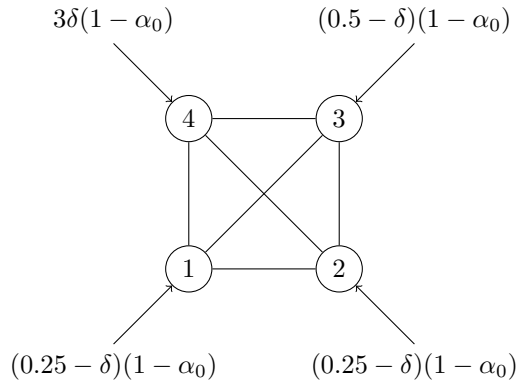
\begin{figure}[htbp]
     \centering
    \begin{tikzpicture}[]
     \node[style={circle,draw}] at (0,0) (1) {$1$};
     \node[style={circle,draw}] at (2,0) (2) {$2$};
     \node[style={circle,draw}] at (2,2) (3) {$3$};
     \node[style={circle,draw}] at (0,2) (4) {$4$};
     \draw[<-] (1) -- (-1,-1)  node[below] {$(0.25-\delta)(1-\alpha_0)$} ;
    \draw[<-] (2) -- (3,-1)  node[below] {$(0.25-\delta)(1-\alpha_0)$} ;
    \draw[<-] (3) -- (3,3)  node[above] {$(0.5-\delta)(1-\alpha_0)$} ;
    \draw[<-] (4) -- (-1,3)  node[above] {$3\delta(1-\alpha_0)$} ;
     \draw (1) -- (2);
     \draw (4) -- (2);
     \draw (1) -- (3);
     \draw (2) -- (3);
     \draw (3) -- (4);
     \draw (4) -- (1);
    \end{tikzpicture}  
   \caption{A matching graph that consists of a complete graph with $4$ nodes.}
   \label{fig:4nodes-delta2}
  \end{figure}

\begin{figure}[t!]
\centering
\includegraphics[width=\columnwidth]{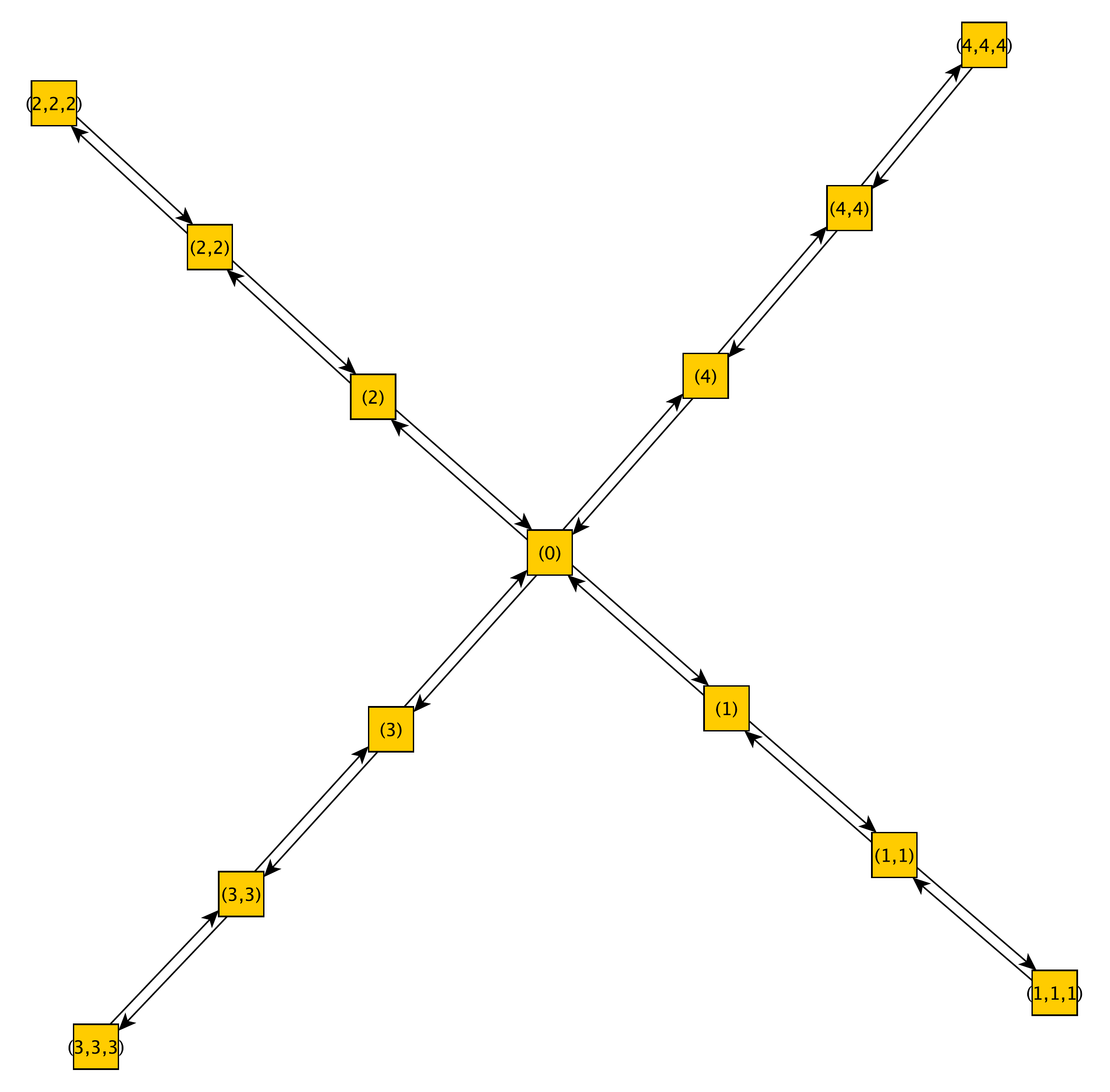}
\caption{The Markov Chain (truncated to words of size 3) derived from the matching model of
Fig.~\ref{fig:4nodes-delta2}.}
\label{fig:4d}
\end{figure}

When we add the edge (1,2) to the matching model of 
Fig.~\ref{fig:4nodes-delta}, it results the matching graph of Fig.~\ref{fig:4nodes-delta2},
which is a complete graph with four nodes. The Markov Chain that describes the dynamics 
of this matching model is represented in Fig.\ref{fig:4d}. In the following result, we provide an 
expression of the mean number of items for this matching model.

\begin{lemma}
For the matching model of 
Fig.~\ref{fig:4nodes-delta2}, the mean number of items is
$$
\frac{2\frac{(0.25-\delta)(0.75+\delta)}{(0.5+2\delta)^2}+\frac{(0.5-\delta)(0.5+\delta)}{(2\delta)^2}+\frac{3\delta(1-3\delta)}{(1-6\delta)^2}}
{1+2\frac{0.25-\delta}{0.5+2\delta}+\frac{0.5-\delta}{2\delta}+\frac{3\delta}{1-6\delta}}
$$
\label{lem:en-4nodes-delta-complete}
\end{lemma}

\begin{proof}
This lemma is the result of Proposition~\ref{prop:finite_sums} where $\Iset=\{\{1\}, \{2\}, \{3\}, \{4\}\}$, $\alpha_1=\alpha_2=0.25-\delta$, $\alpha_3=0.5-\delta$, $\alpha_4=3\delta$, $|\alpha_{\Ecal(\{1\})}|=|\alpha_{\Ecal(\{2\})}|=0.75+\delta$, $|\alpha_{\Ecal(\{3\})}|=0.5+\delta$
and $|\alpha_{\Ecal(\{4\})}|=1-3\delta$.
\end{proof}

We now compare the results of the previous lemmata to determine the conditions under which
a Braess paradox exists. In the following result, we provide 
a necessary and sufficient condition of its existence.

\begin{proposition}
In the matching model of Fig.~\ref{fig:4nodes-delta}, there exists a Braess paradox if and 
only if $\delta\in (0,0.0563)\cup  (0.134,\tfrac{1}{6})$.
\label{prop:quasicomplete-4nodes-delta}
\end{proposition}

\begin{proof}
We aim to study the sign of $\mathbb E[Q]-\mathbb E[\overline Q]$, which is
the same as the sign of
\begin{multline}
\left(\frac{(0.5-2\delta)(0.5+2\delta)}{(4\delta)^2}+\frac{(0.5-\delta)(0.5+\delta)}{(2\delta)^2}+\frac{3\delta(1-3\delta)}{(1-6\delta)^2}\right)\\\left(1+2\frac{0.25-\delta}{(0.5+2\delta)}+\frac{0.5-\delta}{2\delta}+\frac{3\delta}{1-6\delta}\right)-
\\\left(2\frac{(0.25-\delta)(0.75+\delta)}{(0.5+2\delta)^2}+\frac{(0.5-\delta)(0.5+\delta)}{(2\delta)^2}+\frac{3\delta(1-3\delta)}{(1-6\delta)^2}\right)\\
\left(1+\frac{0.5-2\delta}{4\delta}+\frac{0.5-\delta}{2\delta}+\frac{3\delta}{1-6\delta}\right).
\label{eq:eq-diff-alpha0}
\end{multline}

Simplifying this expression, we obtain the following equivalent one:
\begin{multline*}
\frac{54}{\delta^3(0.25+\delta)^2(-1+6\delta)^3} \times\\
\left(
4.52112\cdot 10^{-6}-0.0001808\delta+0.0026222\delta^2-0.0180845\delta^3\right.\\\left.+0.0640914\delta^4-0.133102\delta^5+
0.30556\delta^6-0.8333\delta^7+\delta^8.
\right)
\end{multline*}

We note that $-1+6\delta<0$ when $\delta\in(0,\tfrac{1}{6})$. Thus, since
$$\frac{54}{\delta^3(0.25+\delta)^2(-1+6\delta)^3}$$ is negative when 
$\delta\in(0,\tfrac{1}{6})$, the sign of 
$\mathbb[Q]-\mathbb E[\overline Q]$ is the opposite of 
the sign of the polynomial of degree 8:
\begin{multline*}
4.52112\cdot 10^{-6}-0.0001808\delta+0.0026222\delta^2-0.0180845\delta^3+0.0640914\delta^4\\-0.133102\delta^5+
0.30556\delta^6-0.8333\delta^7+\delta^8.
\end{multline*}

This polynomial has 2 real roots between zero and $\tfrac{1}{6}$, which are 
$0.0563$ and $0.134$. Besides, if $\delta\in (0,0.0563)\cup  (0.134,\tfrac{1}{6})$, the polynomial is positive and negative otherwise. Therefore,
the desired result follows. 
\end{proof}

From the above result, we can conclude that a Braess paradox is achieved when the system
is close to the stability condition. In fact, when $\delta\to 0$, the difference between 
$\alpha_3$ and $\alpha_{\Ecal(3)}$, that is, the difference between the probability of an
arrival to node 3 and the sum of the probability of an arrival to the 
neighbors of node 3, tends to zero. 


In the following result, we quantify the difference between the mean number of items
computed in Lemma~\ref{lem:en-4nodes-delta-complete} and in Lemma~\ref{lem:en-4nodes-delta-quasi}.

\begin{proposition}
When $\delta$ tends to $0$, 
$$
\delta(\mathbb E[\overline Q]-\mathbb E[Q])\to 0.041\wideparen{6}.
$$
\label{prop:4nodes-delta-comparison}
\end{proposition}

\begin{proof}
See Appendix~\ref{proof:prop:4nodes-delta-comparison}
\end{proof}

From the result of Proposition~\ref{prop:quasicomplete-4nodes-delta}, 
the mean number of items of the matching model of Fig.~\ref{fig:4nodes-delta} increases
when we add the edge (1,2) when $\delta$ is sufficiently small. From the above result, we
conclude that, when $\delta\to0,$ the difference $\mathbb E[\overline Q]-\mathbb E[Q]$ 
is unbounded.

\section{Arbitrary Graphs}\label{sec:arbitrary_graph}


An example of the existence of a Braess paradox for matching models was shown in the previous section. This leads to the following questions: (i) could we find a Braess paradox for any arbitrary matching graph? and (ii) if so, what are the conditions for the existence of such a paradox?

In this section, we are going to prove sufficient conditions for the existence or non-existence of a Braess paradox in matching models with an arbitrary matching graphs under an assumption of "saturation" for one independent set of the matching graph. We formalize this assumption as follows:

\begin{assumption}\label{ass:saturated}
We assume that the probabilities $\alpha_1, \cdots,\alpha_n$ are defined as linear combinations of $\delta$, i.e $\alpha_i=a_i+b_i\delta$, with a positive constant ($a_i>0$) and such that $\alpha$ is a probability distribution and satisfy the stability conditions \eqref{eq:stability} for all $\delta\in ]0,\overline{\delta}]$ (with $\overline{\delta}>0$). In addition, we assume that $\left(|\alpha_{\Ecal(\hat{\I})}|-|\alpha_{\hat{\I}}|\right)_a=\sum_{i\in\Ecal(\hat{\I})}a_i - \sum_{i\in\hat{\I}}a_i=0$ for exactly one independent set $\hat{\I}\in\Iset$ and we say that $\hat{\I}$ is saturated.
\end{assumption}
\begin{remark}
The saturated independent set $\hat{\I}$ is defined based on $\Ecal$ and not $\overline{\Ecal}$. This is a consequence of having exactly one saturated independent set. Indeed, if $\hat{\I}$ is saturated for $\overline{\Ecal}$, then it has to be saturated for $\Ecal$ as $|\alpha_{\overline{\Ecal}(\hat{\I})}|\geq |\alpha_{\Ecal(\hat{\I})}|$ (because $\Ecal(\I)\subseteq\overline{\Ecal}(\I)$ for any $\I\in\Iset$) and as $\alpha$ must be positive for all $\delta\in ]0,\overline{\delta}]$. Extending Assumption~\ref{ass:saturated} to multiple saturated independent sets will be discussed at the end of the section.
\end{remark}

We start by giving some notations that will be useful in the remainder of this section. Recall that $\mathcal{G}$ is the original compatibility graph and $\overline{\mathcal{G}}$ is the compatibility graph with the added edge. Let $\I$ be an independent set of $\mathcal G$ and $\Iset$ (resp. $\overline{\Iset}$) be the set of independent sets of $\mathcal G$ (resp. $\overline{\mathcal G}$). 
Furthermore, let $\Iset_{\I}\subseteq\Iset$ be the set of independent sets that are subsets of $\I\in\Iset$, i.e $\Iset_{\I}=\{\tilde{\I}\in\Iset : \tilde{\I}\subseteq\I\}$.
We also define a partition of $\Iset$ based on whether an independent set $\I$ contains the node $i^\ast$ or the node $j^\ast$, i.e $\Iset^\ast=\{\I\in\Iset : i^\ast\in\I \text{ or }j^\ast\in\I\}$ and $\Iset^{-\ast}=\Iset\setminus\Iset^\ast$. 

For all independent sets $\I\in\Iset$, consider an ordered version of $\I$ noted $\I^o=\{i_1,\cdots,i_{|\I|}\}$,  we note $\sigma$ a permutation of its elements, i.e $\I^{\sigma(o)}=\{i_{\sigma(1)},\cdots,i_{\sigma(|\I|)}\}$ and $\mathfrak{S}_{|\I|}$ the set of all permutations of $\llbracket 1,|\I|\rrbracket$. We define
\[
T_{\I^o} = \prod_{k=1}^{|\I|} \frac{\alpha_{i_k}}{|\alpha_{\Ecal(\{i_1,\cdots,i_k\})}|-|\alpha_{\{i_1,\cdots,i_k\}}|}
\]
and $T_{\I}=\sum_{\sigma\in\mathfrak{S}_{|\I|}}T_{\I^{\sigma(o)}}$. We also define
\[
E_{\I^o} = \sum_{l=1}^{|\I|} \frac{ |\alpha_{\Ecal(\{i_1,\cdots,i_l\})}|}{|\alpha_{\Ecal(\{i_1,\cdots,i_l\})}|-|\alpha_{\{i_1,\cdots,i_l\}}|}\prod_{k=1}^{|\I|} \frac{\alpha_{i_k}}{|\alpha_{\Ecal(\{i_1,\cdots,i_k\})}|-|\alpha_{\{i_1,\cdots,i_k\}}|}
\]
and $E_{\I}=\sum_{\sigma\in\mathfrak{S}_{|\I|}}E_{\I^{\sigma(o)}}$. We also note $\overline{T}_{\I}$ and $\overline{E}_{\I}$ for all $\I\in\overline{\Iset}$, where $\Ecal$ is replaced by $\overline{\Ecal}$ in the definition above.

The authors in [21] show that the normalization constant can be written as the sum of terms over the independent sets. In the following result, we show that the stationary distribution can be also written as the sum of terms over the independent sets.

\begin{proposition}\label{prop:finite_sums}
Let $\E[Q]$ be the expected value of $Q$ under the stationary distribution $\pi$. It is defined as an infinite sum over all possible words but can be rewritten as a finite sum over all independent sets, i.e
\[
\E[Q] = \left(1+\sum_{\I\in\Iset}T_{\I}\right)^{-1}\left(\sum_{\I\in\Iset}E_{\I} \right)
\]
\end{proposition}
\begin{proof}
See Appendix~\ref{app:prop:finite_sums}
\end{proof}

We are interested in the sign of the difference in the expected values of $\overline{Q}$ and $Q$, i.e the sign of $\E[\overline{Q}]-\E[Q]$.

Using Proposition~\ref{prop:finite_sums}, we can rewrite the difference in expected values as
\[
\E[\overline{Q}]-\E[Q]=\left(1+\sum_{\I\in\overline{\Iset}}\overline{T}_{\I}\right)^{-1}\left(\sum_{\I\in\overline{\Iset}}\overline{E}_{\I} \right) - \left(1+\sum_{\I\in\Iset}T_{\I}\right)^{-1}\left(\sum_{\I\in\Iset}E_{\I} \right).
\]
Because of Assumption~\ref{ass:saturated}, we know that $\sum_{\I\in\Iset}T_{\I}$ (resp. $\sum_{\I\in\overline{\Iset}}\overline{T}_{\I}$) is positive and thus, the sign of $E[\overline{Q}]-\E[Q]$ is also the sign of
\begin{align}
 &\left(\sum_{\I\in\overline{\Iset}}\overline{E}_{\I}\right)\left(1+\sum_{\I\in\Iset}T_{\I}\right)  - \left(\sum_{\I\in\Iset}E_{\I}\right)\left(1+\sum_{\I\in\overline{\Iset}}\overline{T}_{\I} \right) \nonumber\\
&=\left(\sum_{i\in\overline{\Iset}^\ast}\overline{E}_{\I}\right)\left(1+\sum_{\I\in\Iset}T_{\I}\right) - \left(\sum_{\I\in\Iset^\ast}E_{\I}\right)\left(1+\sum_{\I\in\overline{\Iset}}\overline{T}_{\I} \right) \nonumber\\
&\quad + \left(\sum_{\I\in\Iset^{-\ast}}E_{\I} \right) \left(\sum_{\I\in\Iset^\ast}T_{\I} - \sum_{\I\in\overline{\Iset}^\ast}\overline{T}_{\I}\right)\label{eq:expected_values_diff}
\end{align}
where the equality comes from the fact that $\overline{\Iset}^{-\ast}=\Iset^{-\ast}$, $\overline{\Ecal}(\I)=\Ecal(\I)$ for all $\I\in\Iset^{-\ast}$. Indeed, because of that, $\overline{E}_{\I}=E_{\I}$ and $\overline{T}_{\I}=T_{\I}$ for all $\I\in\Iset^{-\ast}$.

In order to ease the following proofs, we add some notations for the difference in expected values. Let $A_1$, $A_2$, $B_1$, $B_2$, $C_1$ and $C_2$ be equal to the terms in parenthesis in \eqref{eq:expected_values_diff} from left to right respectively. We also define $A=A_1 \cdot A_2$, $B=B_1 \cdot B_2$ and $C=C_1\cdot C_2$ such as $\eqref{eq:expected_values_diff}$ can be rewritten as $A-B+C=A_1\cdot A_2 - B_1 \cdot B_2 + C_1 \cdot C_2$. Each of these terms can be written as a rational fraction of $\delta$ because of Assumption~\ref{ass:saturated}. We are interested in their limit when $\delta$ tends to zero. For that purpose, we introduce the following two lemmas that will help us in the proof of the main result. 

\begin{lemma}\label{lem:pos_cst_in_sums}
Given Assumption~\ref{ass:saturated}, the terms $A_1$, $A_2$, $B_1$, $B_2$ and $C_1$ can be written as a rational fraction of $\delta$ such that the polynomial at the numerator has a positive constant term.
\end{lemma}
\begin{proof}
See Appendix~\ref{app:lem:pos_cst_in_sums}
\end{proof}

\begin{lemma}\label{lem:pos_cst_in_sums_C}
Given Assumption~\ref{ass:saturated}, if $\hat{\I}\notin \bigcup_{\I\in\Iset^\ast}\Iset_{\I}$, then $C_2$ can be written as a rational fraction of $\delta$ such that the polynomial at the numerator has a positive constant term.
\end{lemma}
\begin{proof}
See Appendix~\ref{app:lem:pos_cst_in_sums_C}
\end{proof}

We are now ready to present the main result of this paper about the existence or non-existence of a Braess paradox for matching models under Assumption~\ref{ass:saturated}.

\begin{theorem}\label{thm:saturated_braess_paradox}
Given Assumption~\ref{ass:saturated}, if $\hat{\I}\in\Iset^{-\ast}\setminus \left(\bigcup_{\tilde{\I}\in\Iset^\ast}\Iset_{\tilde{\I}}\right)$, then there exists a Braess paradox for $\delta$ sufficiently small. If $\hat{\I}\in\Iset^\ast $, then there does not exist a Braess paradox for $\delta$ sufficiently small.
\end{theorem}
\begin{proof}
The existence or non-existence of the Braess paradox depends on the sign of $\E[\overline{Q}]-\E[Q]$. We start by rewriting the difference in expected values as in \eqref{eq:expected_values_diff}. Then, we put every terms on the same denominator using a similar approach as in Lemma~\ref{lem:pos_cst_in_sums} and Lemma~\ref{lem:pos_cst_in_sums_C}. For the term $A$, the common denominator between all the terms in $A_1$ and in $A_2$ is equal to
\begin{align}
&\prod_{\I\in\overline{\Iset}^\ast_+\setminus\overline{\Iset}^\ast}\left(|\alpha_{\Ecal(\I)}|-|\alpha_{\I}|\right)^2\prod_{\I\in\overline{\Iset}^\ast}\left(|\alpha_{\overline{\Ecal}(\I)}|-|\alpha_{\I}|\right)^2  \prod_{\I\in\Iset}\left(|\alpha_{\Ecal(\I)}|-|\alpha_{\I}|\right)
\label{eq:A_denom}
\end{align}
For the term $B$, the common denominator between all the terms in $B_1$ and in $B_2$ is equal to
\begin{align}
&\prod_{\I\in\Iset^\ast_+}\left(|\alpha_{\Ecal(\I)}|-|\alpha_{\I}|\right)^2\prod_{\I\in\overline{\Iset}^{\ast}}\left(|\alpha_{\overline{\Ecal}(\I)}|-|\alpha_{\I}|\right)  \prod_{\I\in\overline{\Iset}^{-\ast}}\left(|\alpha_{\Ecal(\I)}|-|\alpha_{\I}|\right)
\label{eq:B_denom}
\end{align}
For the term $C$, the common denominator between all the terms in $C_1$ and in $C_2$ is equal to
\begin{align}
&\prod_{\I\in\Iset^{-\ast}}\left(|\alpha_{\Ecal(\I)}|-|\alpha_{\I}|\right)^2 \prod_{\I\in\Iset^\ast_+\setminus\Iset^\ast}\left(|\alpha_{\Ecal(\I)}|-|\alpha_{\I}|\right)\prod_{\I\in\Iset^\ast}\left(|\alpha_{\Ecal(\I)}|-|\alpha_{\I}|\right)\nonumber\\
&\times\prod_{\I\in\overline{\Iset}^\ast}\left(|\alpha_{\overline{\Ecal}(\I)}|-|\alpha_{\I}|\right)
\label{eq:C_denom}
\end{align}
Therefore the common denominator for \eqref{eq:expected_values_diff} is equal to
\begin{align}
&\prod_{\I\in \Iset^\ast_+\setminus\Iset^\ast}\left(|\alpha_{\Ecal(\I)}|-|\alpha_{\I}|\right)^3 \prod_{\I\in\Iset^\ast\cup\left(\Iset^{-\ast}\setminus\Iset^\ast_+\right)}\left(|\alpha_{\Ecal(\I)}|-|\alpha_{\I}|\right)^2 \nonumber\\
&\times\prod_{\I\in\overline{\Iset}^\ast}\left(|\alpha_{\overline{\Ecal}(\I)}|-|\alpha_{\I}|\right)^2
\label{eq:expected_diff_denom}
\end{align}
because $\Iset^{-\ast}=\overline{\Iset}^{-\ast}$, $\left(\Iset^{\ast}_+\setminus\Iset^\ast\right)\subseteq \Iset^{-\ast}$ and $\Iset^{\ast}_+\setminus\Iset^\ast = \overline{\Iset}^{\ast}_+\setminus\overline{\Iset}^\ast$.

Using Assumption~\ref{ass:saturated}, we know that there is exactly one independent set $\hat{\I}\in\Iset$ such that $\left(|\alpha_{\Ecal(\hat{\I})}|-|\alpha_{\hat{\I}}|\right)=\left(\sum_{i\in\Ecal(\hat{\I})}b_i-\sum_{i\in\hat{\I}}b_i \right)\delta$ which tends to $0^+$ when $\delta$ tends to $0^+$ (because $\sum_{i\in\Ecal(\hat{\I})}b_i-\sum_{i\in\hat{\I}}b_i>0$ as $\alpha$ satisfy the stability condition). 
For all the other terms in \eqref{eq:expected_diff_denom} that depends on a different independent set than $\hat{\I}$, they will tend to a positive constant because of Assumption~\ref{ass:saturated}.
Thus, the denominator will tends to $0^+$ when $\delta$ tends to $0^+$ and the only concern now is if the polynomial in $\delta$ at the numerator has a constant and what is its sign.

In order to get the denominator in \eqref{eq:expected_diff_denom}, the numerator of term $A$ will be multiplied by
\begin{align}
\prod_{\I\in\Iset^\ast\cup\left(\Iset^{-\ast}\setminus\Iset^\ast_+\right)}\left(|\alpha_{\Ecal(\I)}|-|\alpha_{\I}|\right), 
\label{eq:A_num_mul}
\end{align}
the numerator of term $B$ will be multiplied by
\begin{align}
\prod_{\I\in \Iset^{-\ast}\setminus\Iset^\ast_+}\left(|\alpha_{\Ecal(\I)}|-|\alpha_{\I}|\right) \prod_{\I\in\overline{\Iset}^\ast}\left(|\alpha_{\overline{\Ecal}(\I)}|-|\alpha_{\I}|\right)
\label{eq:B_num_mul}
\end{align}
and the numerator of term $C$ will be multiplied by
\begin{align}
\prod_{\I\in\Iset^\ast}\left(|\alpha_{\Ecal(\I)}|-|\alpha_{\I}|\right) \prod_{\I\in\overline{\Iset}^\ast}\left(|\alpha_{\overline{\Ecal}(\I)}|-|\alpha_{\I}|\right). 
\label{eq:C_num_mul}
\end{align}

If $\hat{\I}\in\Iset^{-\ast}\setminus\Iset^\ast_+$, then \eqref{eq:A_num_mul} and \eqref{eq:B_num_mul} will have a factor with no constant (the one related to $\hat{\I}$), whereas \eqref{eq:C_num_mul} will have a positive constant because of Assumption~\ref{ass:saturated}. Therefore, the value of the constant at the numerator of \eqref{eq:expected_values_diff} can only depend on the term $C$ which has a positive constant because of Lemma~\ref{lem:pos_cst_in_sums} and Lemma~\ref{lem:pos_cst_in_sums_C}. In conclusion, the denominator of \eqref{eq:expected_values_diff} tends to $0^+$ when $\delta$ tends to $0^+$ and its numerator tends to a positive constant which means that for $\delta$ sufficiently small, \eqref{eq:expected_values_diff} is positive and so does $\E[\overline{Q}]-\E[Q]$. 

If $\hat{\I}\in\Iset^{\ast}$, then \eqref{eq:A_num_mul} and \eqref{eq:C_num_mul} will have a factor with no constant (the one related to $\hat{\I}$), whereas \eqref{eq:B_num_mul} will have a positive constant because of Assumption~\ref{ass:saturated}. Therefore, the value of the constant at the numerator of \eqref{eq:expected_values_diff} can only depend on the term $B$ which has a positive constant because of Lemma~\ref{lem:pos_cst_in_sums}. In conclusion, the denominator of \eqref{eq:expected_values_diff} tends to $0^+$ when $\delta$ tends to $0^+$ and its numerator tends to a negative constant which means that for $\delta$ sufficiently small, \eqref{eq:expected_values_diff} is negative and so does $\E[\overline{Q}]-\E[Q]$. 
\end{proof}

\begin{figure}[htbp]
     \centering
    \begin{tikzpicture}[]
     \node[style={circle,draw}] at (0,0) (1) {$1$};
     \node[style={circle,draw}] at (2,0) (2) {$2$};
     \node[style={circle,draw}] at (2,2) (3) {$3$};
     \node[style={circle,draw}] at (0,2) (4) {$4$};
     \node[style={circle,draw}] at (-1,1) (5) {$5$};
     \node[style={circle,draw}] at (1,3) (6) {$6$};
     \draw[<-] (1) -- (-1,-1)  node[below] {$(0.1+\delta)(1-\alpha_0)$} ;
    \draw[<-] (2) -- (3,-1)  node[below] {$(0.1+\delta)(1-\alpha_0)$} ;
    \draw[<-] (3) -- (3,3)  node[above] {$(0.1+\delta)(1-\alpha_0)$} ;
    \draw[<-] (4) -- (-1,3)  node[above] {$(0.25+\delta)(1-\alpha_0)$} ;
    \draw[<-] (5) -- (-2,1.5)  node[above] {$(0.35-5\delta)(1-\alpha_0)$} ;
    \draw[<-] (6) -- (1,4)  node[above] {$(0.1+\delta)(1-\alpha_0)$} ;
     \draw (4) -- (2);
     \draw (1) -- (3);
     \draw (2) -- (3);
     \draw (3) -- (4);
     \draw (4) -- (1);
     \draw (5) -- (1);
     \draw (5) -- (4);
     \draw (6) -- (1);
     \draw (6) -- (2);
     \draw (6) -- (3);
     \draw (6) -- (4);
    \end{tikzpicture}  
   \caption{A matching graph that consists of a quasi-complete graph with $5$ nodes plus one node connected to two of them.}
   \label{fig:quasi_complete_5nodes_plus1}
  \end{figure}
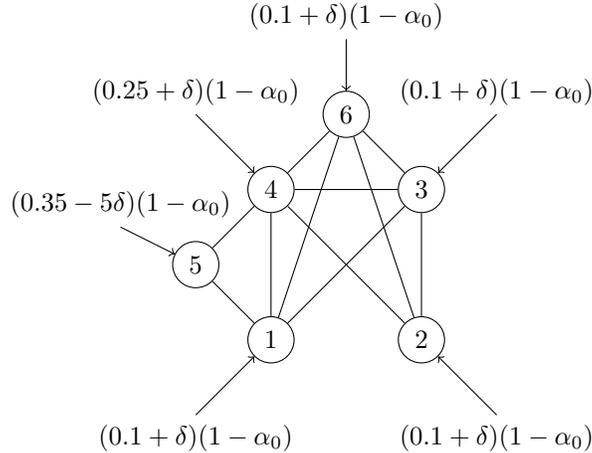

Theorem~\ref{thm:saturated_braess_paradox} covers a lot of independent sets but if the saturated independent set does not contain $i^\ast$ or $j^\ast$ and does not have them both in its neighbors (i.e $\hat{\I}\in\Iset^{-\ast}\cap\left(\bigcup_{\I\in\Iset^\ast}\Iset_{\I}\right)$), then we did not succeed in proving the existence or non-existence of the Braess paradox. However, we have made numerical experiments which suggest that a Braess paradox exists in that case as well. Indeed, consider the matching model with a matching graph $\mathcal{G}$ as shown in Fig.~\ref{fig:quasi_complete_5nodes_plus1} and let $(1, 2)$ be the added edge in $\overline{\mathcal{G}}$.
Using $\alpha$ as defined in the figure there is only one saturated independent set which is $\hat{\I}=\{5\}$ and is not considered in Theorem~\ref{thm:saturated_braess_paradox} ($\{5\}\in\Iset^{-\ast}$ and $\{5\}\subseteq \{2,5\}\in\Iset^{\ast}$). Using Proposition~\ref{prop:finite_sums}, we computed the difference in expected values for $\delta=0.001$ and we obtained $\E[\overline{Q}]-\E[Q]=0.0903021657941>0$ which means that there is a Braess paradox for this model.

To conclude this section, we want to discuss a part of Assumption~\ref{ass:saturated} which is the fact that there is exactly one saturated independent set. A possible extension of our result would be to allow multiple saturated independent sets. Let $\hat{\Iset}$ be the set of saturated independent sets. If all the saturated sets belong to the same "category", i.e $\hat{\Iset}\subseteq \Iset^{\ast}$ or $\hat{\Iset}\subseteq \Iset^{-\ast}\setminus\left(\bigcup_{\I\in\Iset^{\ast}}\Iset_{\I}\right)$, then the same arguments of Theorem~\ref{thm:saturated_braess_paradox} can be used to show the existence or non-existence of the paradox. However, when all the saturated independent sets do not belong to the same "category",  the existence of a Braess paradox remains an open question.

\section{Extensions}\label{sec:extensions}

\newcommand{\ve}{{\mathcal{V}}}
\newcommand{\mac}{{\mathcal{M}}}
\newcommand{\Ga}{{\mathcal{E}}}
\newcommand{\Gr}{{\mathcal{G}}}

We now address a simple question on the size of a matching graph which can exhibit a Braess paradox. In section \ref{sec:quasi_complete}
we have presented such an example: a complete graph with $4$ nodes and the same graph with one edge deleted. 
This example shows that it is possible to obtain a paradox with a $4$ nodes graph. The following property states that 
it is not possible to have a paradox with a graph  with 
a smaller number of vertices. 

\begin{proposition} 
Matching Graphs  with $2$ or $3$ nodes do not exhibit Braess paradox. 
 \end{proposition} 
\begin{proof}
Indeed, under our assumptions on the arrivals, bipartite matching graphs are 
associated to unstable Markov chains. Therefore one must restrict ourselves to matching graphs which are not bipartite.
The smallest matching graph associated to a stable stochastic model  is  the complete graph with $3$ nodes. But if 
 we remove one edge to this complete graph, the resulting graph is a path of length $3$ which is again bipartite. Therefore such a paradox 
 does not exist as its associated stochastic models is not stable. 
\end{proof}

We now prove a constructive method to extend the number of nodes in a matching graph with paradox 
while keeping the paradox for a similar distribution of arrivals. 
We show in Theorem \ref{fin} how one can obtain a new matching graph with a paradox from an old one also with a paradox. The new matching graph has one more node and several more edges. 
Combining these results and the example of Section \ref{sec:quasi_complete}, we obtain the following proposition:

\begin{proposition} 
For all number of nodes $n>3$, there exists Matching Graphs with $n$ nodes which  exhibit Braess paradox. 
 \end{proposition} 
\begin{proof}
Indeed, the example of Section~\ref{sec:quasi_complete} shows that a paradox exist for the complete graph with $4$ nodes and Theorem~\ref{fin} states that if a paradox
exists for a graph with $n$ nodes, it also exists for a graph with $n+1$ nodes. By induction, the proposition is proved. 
\end{proof}

We first show how to derive two matching graphs which have the same expected size of words in steady-state. 
This construction is based on lumpability. For a definition of ordinary lumpability, see Ref. \cite{KeSn60}. 
Note that this construction does not rely on the saturation property studied in the previous section. 
We will elaborate more on this topic at the end of this section. 

\begin{figure} [hbtp]
\begin{center}
    \begin{tikzpicture}[]
     \node[style={circle,draw}] at (-4,2) (1) {$y$};
     \node[style={circle,draw}] at (-3,2) (2) {$z$};
     \node[style={circle,draw}] at (-4,0) (3) {$$};
     \node[style={circle,draw}] at (-3.5,0) (4) {$$};
     \node[style={circle,draw}] at (-3,0) (5) {$$};
     \node[style={circle,draw}] at (-4,-0.5) (6) {$$};
     \node[style={circle,draw}] at (-3.5,-0.5) (7) {$$};
     \node[style={circle,draw}] at (-4,-1.5) (8) {$$};
     \node[style={circle,draw}] at (-3.5,-1) (9) {$$};
     \node[style={circle,draw}] at (-3.5,-1.5) (10) {$$};
     \draw (5) -- (1);
     \draw (5) -- (2);
     \draw (4) -- (2);
     \draw (1) -- (3);
     \draw (2) -- (3);
     \draw (3) -- (4);
     \draw (4) -- (1);
    \draw (6) -- (4);
     \draw (7) -- (5);
     \draw (7) -- (4);
    \draw (9) -- (7);
     \draw (8) -- (9);
     \draw (7) -- (6);

     \draw (9) -- (10);
     
      \node[style={circle,draw}] at (0.5,2) (11) {$x$};
     \node[style={circle,draw}] at (0,0) (13) {$$};
     \node[style={circle,draw}] at (0.5,0) (14) {$$};
     \node[style={circle,draw}] at (1,0) (15) {$$};
     \node[style={circle,draw}] at (0,-0.5) (16) {$$};
     \node[style={circle,draw}] at (0.5,-0.5) (17) {$$};
     \node[style={circle,draw}] at (0,-1.5) (18) {$$};
     \node[style={circle,draw}] at (0.5,-1) (19) {$$};
     \node[style={circle,draw}] at (0.5,-1.5) (20) {$$};
     \draw (15) -- (11);
     \draw (11) -- (13);
     \draw (13) -- (14);
     \draw (14) -- (11);
    \draw (16) -- (14);
     \draw (17) -- (15);
     \draw (17) -- (14);
    \draw (19) -- (17);
     \draw (18) -- (19);
     \draw (17) -- (16);

     \draw (19) -- (20);

     \end{tikzpicture}  
\end{center}
\caption{Decomposition matching graph (left), Aggregated matching graph (right). \label{matchlump}}
\end{figure}
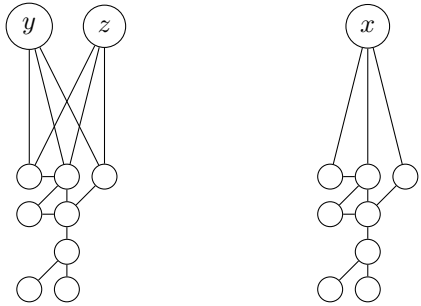

We first consider an arbitrary matching graph and an arbitrary node $x$. 
Let us denote by $\Gr_x$ this graph  (see right part of Fig. \ref{matchlump}). 
Let $W_x$ denote the Markov chain associated with matching $\Gr_x$. 

\begin{definition} [Decomposition matching  and Aggregated matching] \label{agregmatching} 
We define a new matching by an decomposition of $x$ into two nodes $y$ and $z$. 
The decomposition is defined by:
\begin{itemize}
\item $\Ga (y) = \Ga (z) = \Ga(x) $
\item $\alpha_y >0$ 
\item $\alpha_z >0$ 
\item $\alpha_y + \alpha_z = \alpha_x$.
\end{itemize}  

Let $\Gr_{yz}$ be the matching graph where $x$ is decomposed into $y$ and $z$.
$\Gr_{yz}$ will be denoted as the decomposition matching while $\Gr_{x}$ is the aggregated matching. 
$W_{yz}$ is the Markov chain associated with $\Gr_{yz}$. 
\end{definition} 

\begin{proposition} If the aggregated matching is associated to a stable Markov chain, the same property holds for the decomposition matching.
\end{proposition} 

Note that by construction $y \notin \Ga(z) $ and  $z \notin \Ga(y)$. 
We use an event based representation of the chain \cite{Mass87}. Let $e_t$ be the event associated with the arrival of letter t. The probability 
of $e_t$ is $\alpha_t$. $e_t $ is a function from $\V $ to  $\V $.

\begin{theorem} \label{lump} 
$M_{yz}$ is ordinary lumpable according the following partition of the states :
\begin{itemize}
\item Consider a word $w$ which contains a positive number of $y$ and $z$. 
 The macro state which contains  word $w$ contains all the words obtained by exchanging 
letters $y$ and $z$ in $w$. An abstract representation of this set 
is obtained by  replacing  $y$ and $z$ by   letter $x$. 
For instance the state (or word) $(a,a,y,b,a,z,y)$ of $W_{yz}$ is assigned 
to macro-state noted  $(a,a,x,b,a,x,x)$. To obtain all the states which belong to macro state $(a,a,x,b,a,x,x)$, one must replace all the $x$ by a
$y$ or a $z$. Thus, there are $8$ states in macro state $(a,a,x,b,a,x,x)$. We  denote by $\mac(w)$ the  macro state containing $w$. 
\item
If  word $w$ does not contain neither $x$ nor $y$,  then the macro-state is a singleton: $\mac(w)$ only contains $w$. 
\end{itemize}

Furthermore, the lumped Markov chain obtained  from $W_{yz} $ and the described partition 
is the Markov chain associated with matching $\Gr_x$.  
\end{theorem} 
\begin{proof} Let $w$ be an arbitrary state of $W_{yz}$. We will prove that for all $w1$ in $\mac(w)$ then we must have
$e_t (w1) \in \mac(e_t(w))$.  

Remark that if word $w$ contains neither $y$ nor  $z$, the property is trivial. 
Thus, 
without loss of generality, we can assume that word $w$ contains at least one $y$ or one $z$.   

Let $w1$ be an arbitrary state in $\mac(w)$ and $t$ be an arbitrary letter distinct from $y$ and $z$. 
At the arrival of the letter, a matching may happen (or not) with word $w$. 

\begin{itemize}
\item If the matching does not occur, it means that letter $t$ does not match with any letters in $w$. Remark that, as $w1$ is in $\mac(w)$
it contains the same letters than $w$ at the same place except the $y$ and the $z$ which may be exchanged. Remember that $y$ and $z$ 
have the same neighborhood in the matching graph $\Gr_{yz}$. Therefore, changing the configuration of $y$ and $z$ does not change 
the matching with $t$. Combining both arguments, we obtain that $t$ does not match with $w1$. For both words $w$  and $w1$, letter $t$ 
is appended at the end of the word. Thus, we have in that case: $e_t (w1) \in \mac(e_t(w))$. 

\item Consider now that the matching happens between letter $t$ and a letter in $w$. Clearly this letter is also deleted in $w1$. Indeed, the letters
in $w1$ are also in $w$, at the same position, except the $y$ and the $z$ which may be exchanged and which deleted by the same letters
as $\Ga(y) = \Ga(z)$.  Thus $e_t (w1) \in \mac(e_t(w))$ also holds. 
\end{itemize}
Finally assume that letter $t$ is $y$ or $z$. Note that by construction we have $\Ga(y) = \Ga(z)$,  $z \notin \Ga(y) $ and $y \notin \Ga(z) $. 
Therefore the arrival of a $y$ or a $z$ has the 
same effect  (i.e. deletion or appending letter $t$ at the end of the word).. Therefore 
\[
\mac(e_y (w))  = \mac(e_z (w))  = \mac(e_y (w1))  = \mac(e_z (w1)) 
\]
Thus the arrival of $y$ and $z$ can be combine to obtain the same transition and we can add up the transition probabilities. 

To conclude,  we have proved that for all word $w$ of $W_{yz}$ and  for all $w1$ in $\mac(w)$ and for all letter $t$,  we  have
$e_t (w1) \in \mac(e_t(w))$.  
Thus, 
\[
\mac(e_t (w))  = \mac(e_t (w1))
\]
Remember that the macro states used to define the lumpability partition are the sets $\mac(w)$.  
As event $e_t$ have a constant probability (i.e. which is not state dependent), the 
probability $\sum_{j \in A_v} Pr(w,j)$ does not depend on initial state $w$. 
Thus the Markov chain is lumpable for this partition. 

Let us now prove that this is the Markov chain associated with matching graph $\Gr_1$. As $\Ga(x)=\Ga(y)=\Ga(z)$ 
and the remaining matching graphs are the same, the state space of the both Markov chains are identical. 
The fact that transitions are the same follow from the property 
$ \mac(e_t (w))  = \mac(e_t (w1)) $ and because $\alpha_x = \alpha_y + \alpha_z$ while other probabilities do not change. 
\end{proof} 

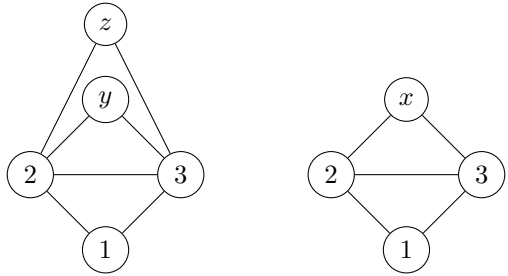
\begin{figure} [hbtp] \label{exa} 
\begin{center}
    \begin{tikzpicture}[]
     \node[style={circle,draw}] at (1,0) (1) {$1$};
     \node[style={circle,draw}] at (0,1) (2) {$2$};
     \node[style={circle,draw}] at (2,1) (3) {$3$};

     \node[style={circle,draw}] at (1,2) (4) {$y$};
     \node[style={circle,draw}] at (1,3) (5) {$z$};
     \draw (2) -- (1);
     \draw (2) -- (3);
     \draw (1) -- (3);
     \draw (2) -- (4);
     \draw (2) -- (5);
     \draw (3) -- (4);
     \draw (3) -- (5);

    \node[style={circle,draw}] at (5,0) (6) {$1$};
     \node[style={circle,draw}] at (4,1) (7) {$2$};
     \node[style={circle,draw}] at (6,1) (8) {$3$};

     \node[style={circle,draw}] at (5,2) (9) {$x$};
     \draw (6) -- (7);
     \draw (6) -- (8);
     \draw (7) -- (8);
     \draw (7) -- (9);
     \draw (8) -- (9);
    \end{tikzpicture}  
\end{center}
\caption{Decomposition matching graph (left), Aggregated matching graph (right). \label{examplelump}}
\end{figure}

For instance we consider the agregated matching in the right part Fig. \ref{exa}. Node $x$ is decomposed in nodes $y$ and $z$ to build 
the decomposition matching graph (left part of the same figure). 

\begin{figure}[t!] 
\centering
\includegraphics[width=\columnwidth]{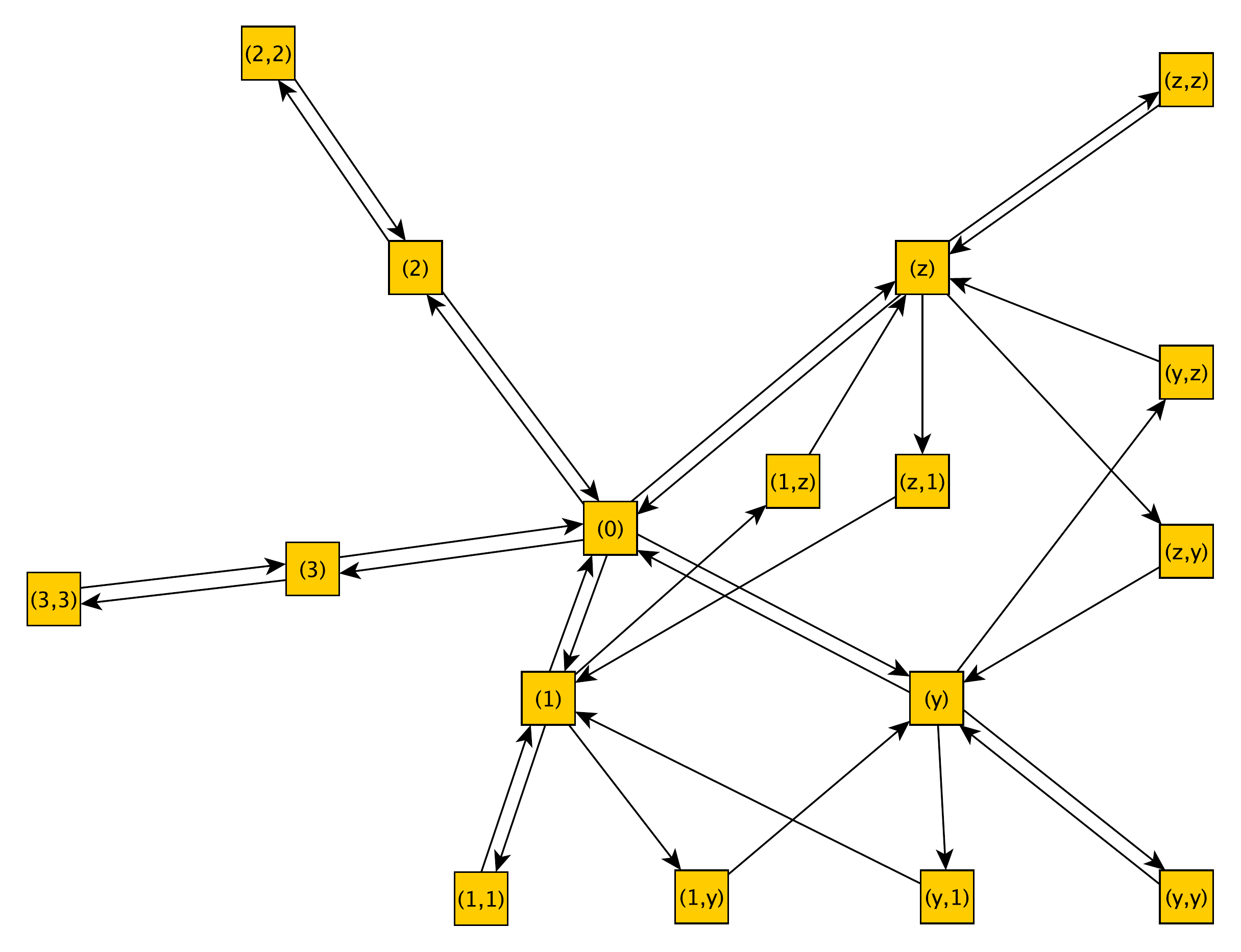}
\caption{Markov chain for the decomposition marking graph restricted to the states with less than 2 letters. }
\label{decom} 
\end{figure}

\begin{figure}[t!]  
\centering
\includegraphics[width=\columnwidth]{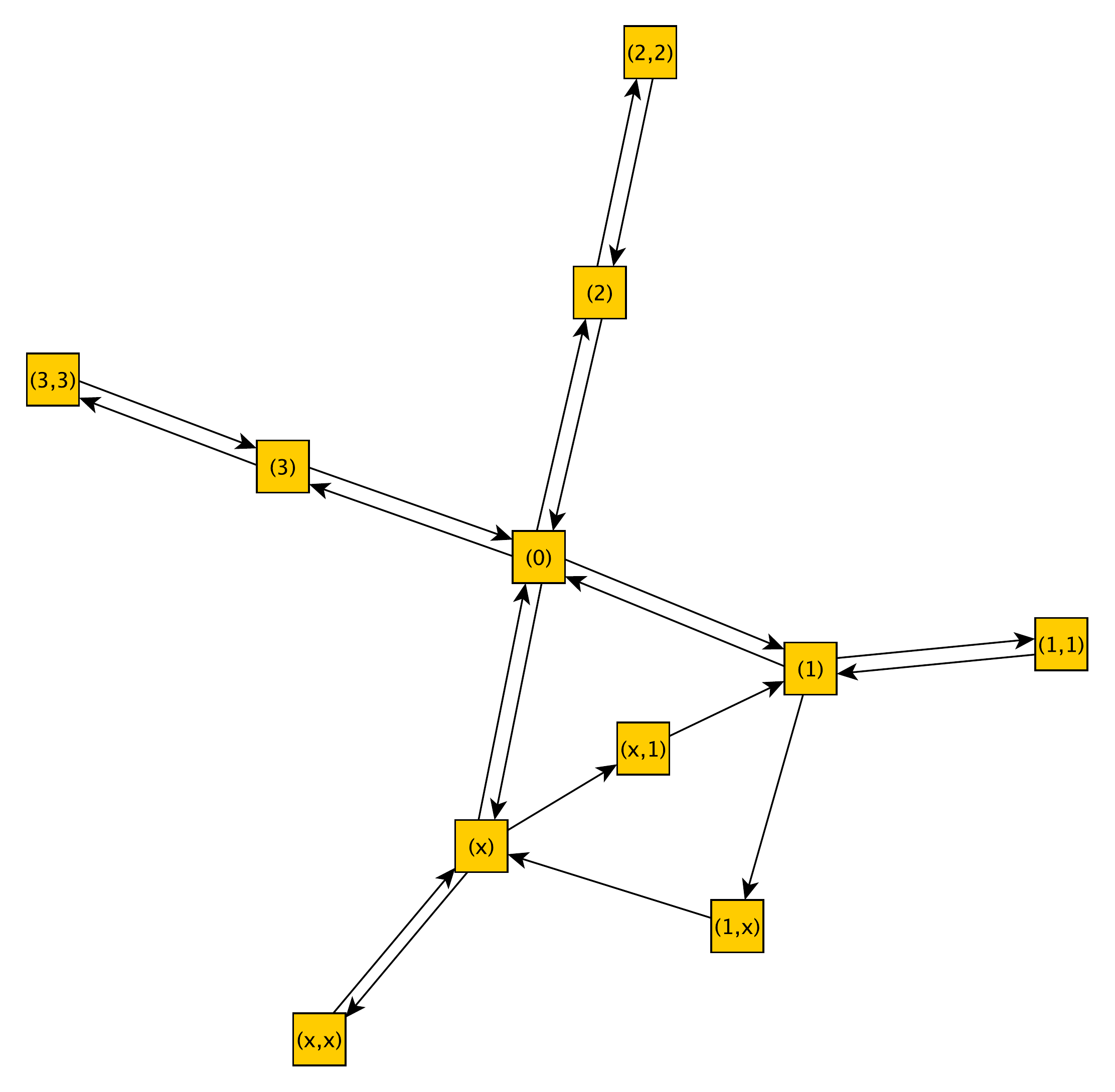}
\caption{Markov chain for the agregated marking graph restricted to the states with less than 2 letters.}
\label{agreg}
\end{figure}

We have built the Markov chains associated to these two matching graphs. They are depicted in Fig. \ref{decom} 
and \ref{agreg}. Of course as these chains is infinite we only represent few states. We have chosen to draw the words with 
 $2$ letters or less.  

\begin{lemma} \label{equallump} 
Let $\E[Q_x]$ (resp. $\E[Q_{yz}]$)  be the average size of a word of $W_x$ (resp. $W_{yz}$). We have 
  $\E[Q_x] = \E[Q_{yz}]$. 
\end{lemma}
\begin{proof} Let $\pi_x $ the steady-state distribution of $W_x$ and $\pi_{yz} $ the steady-state distribution of $W_{yz}$. Similarly let 
$\ve_x^*$ (resp. $\ve_{yz}^*$) be the state space of $W_x$ (resp. $W_{yz}$). Remember that $ | w | $ is the size of the word $w$. 
By definition:
\[
\E[Q_{yz}] =  \sum_{w \in \ve_{yz}^*} \pi_{yz}(w) ~| w | 
\]
We decompose the summation according to the macro state $\mac()$. 
\[
\E[Q_{yz}] =  \sum_{v \in \ve_x^*} \sum_{w \in \mac(v)} \pi_{yz} (w) ~| w | 
\]
For all states $w$ in $\mac(v) $, we clearly have $| w | = | v | $. 
Thus, 
\[
\E[Q_{y,z}] =  \sum_{v \in \ve_x^*} ~| v |  \sum_{w \in \mac(v)} \pi_{yz} (w) 
\]
As the chain is lumpable, we have $\sum_{w \in \mac(v)} \pi_{yz}(w)  = \pi_{x} (v)$. 
Thus, 
\[
\E[Q_{yz}] =  \sum_{v \in \ve_x^*} ~| v |   \pi_x(v) = \E[Q_x]. 
\]
\end{proof}

\begin{theorem} \label{fin} 
Let $\Gr_x$ be a matching graph, associated with a distribution of arrivals $(\alpha_i)$ 
such that there exists a paradox: adding edge $(u,v)$ in $\Gr_x$ (to obtain a new matching graph $\overline{\Gr_x}$) 
results in a larger expected number of letters.  

We consider an arbitrary node $x$ distinct from both $u$ and $v$. We build a new matching graph by a decomposition of node $x$ into nodes $y$ 
and $z$ as mentioned in Definition \ref{agregmatching}. Let 
$\Gr_{yz}$ be the decomposition matching graph we have obtained. 
Furthermore, we divide $\alpha_x$ into two positive parts $\beta_y$ and $\beta_z$ not necessarily equal (i.e. $\alpha_x = \beta_y + \beta_z$).
The others arrival probability are kept unchanged (i.e. $ \beta_i = \alpha_i $ for all $i \neq x$). 
Then, the paradox exists for matching graph $\Gr_{yz}$ associated with distribution of arrivals $(\beta_i)$. 
\end{theorem} 
\begin{proof} 
Let $\overline{\Gr_{yz}}$ be the graph obtained from $\Gr_{yz}$ by doing the decomposition. 
We know from the assumption about the paradox between $\Gr_x$ and $\overline{\Gr_x}$ that:
\[ 
\E[Q_x]  < \E[\overline{Q_x}] 
\]
And from Lemma \ref{equallump}, we have:
\[
\E[Q_x] = \E[Q_{yz}] \text{ and } \E[\overline{Q_x}] = \E[\overline{Q_{yz}}]
\]
As $x$ is neither $u$, nor $v$,  $\overline{\Gr_{yz}} $ is also the matching graph obtained from $\Gr_{yz}$ by adding edge $(u,v)$. 
Combining these results, we have a paradox between $\Gr_{yz}$ and $\overline{\Gr_{yz}}$: 
\[ 
\E[Q_{yz}]  < \E[\overline{Q_{yz}}] 
\]
\end{proof}
We have built by decomposition a new matching graph with one more node and the paradox exists for this new graph if it exits for the first one.

We can iterate with this process building a larger graph at each step. Then one can obtain matching graphs with  paradox of arbitrary size (larger than 4, the size of the smallest graph we have considered in Section \ref{sec:quasi_complete}). The distribution of arrivals on the decomposition matching is derived
from the distribution on the aggregated (i.e. initial) matching. But we do not make any assumptions, except the stability, for all models. This construction gives us the opportunity to observe Braess paradox which are not based on the saturation condition as established in the next proposition. 

\begin{proposition}
Assume that there exists a matching graph $\Gr$ with $n$ nodes and  with a distribution of arrivals which does not follow the saturation condition 
and such that a paradox exists. Then the construction of Theorem \ref{fin} proves that such a paradox also exists for graphs with more than $n$ vertices. 
\end{proposition}

\section{Conclusions}\label{sec:conclusion}

When an edge is added
to the compatibility graph of a matching model, one might think that the mean number of items in the system decreases 
since the number of potential matchings increases. In this work, we have shown that this is not
always the case when the matching policy is FCFS, that is, when an incoming item is 
matched with the oldest of its compatible items. Therefore, a Braess paradox exists for the
family of matching models under consideration in this article.
  
We have studied the existence of a Braess paradox in a matching model whose compatibility
graph is a quasicomplete graph with four nodes. We have established necessary and sufficient 
conditions on the values of the arrivals such that, when we add an edge, the mean number
of items increases. We have also shown that there exists values of the arrivals such that,
when the paradox exists, the difference between the mean number of items of both models
tends to infinity.

We have considered a matching model with an arbitrary graph and we have studied the 
existence of a Braess paradox for this model. First, we have shown that the mean number of
items in the system can be written as a finite sum over all independent sets. This result
has allowed us to show the main contribution of this paper, which gives sufficient conditions
for the existence of a Braess paradox and for the non-existence. Regarding the former result, 
we have shown that a Braess paradox exists when an independent set, that does not 
contain any of the nodes of the edge we add but has them as neighbors, is in saturation. For the latter result, 
we have shown that a Braess paradox does not exist when an independent set that 
is in saturation contains one of the nodes of the edge we add. 

We have also presented an interesting property of the mean number of items of the system
when we add a class of items in the system. For this case, we have shown that the arrivals can
be set in a way such that the mean number of items remains unchanged when we add a class
of items. This implies that, from a matching model where a Braess paradox exists, we can
build an infinite number of matching models such that the Braess paradox exists as well.

For future work, we aim to study analytically the existence of a Braess paradox when the independent set that is saturated does not 
contain any of the nodes of the edge we add and does not have them as neighbors. We also aim to relax the assumption of having exactly one saturated independent set to multiple ones.
Another interesting extension of this work is
to analyze the existence of a Braess paradox for other matching policies. We are also
interested in investigating the existence of a Braess paradox for matching models whose 
compatibility graph is bipartite.


\bibliographystyle{plain}
\bibliography{braessmatching}

\appendix

\section{Proof of Proposition~\ref{prop:4nodes-delta-comparison}}
\label{proof:prop:4nodes-delta-comparison}

If the mean number of customers of the
quasicomplete graph is $\mathbb E[Q]$ and of the complete graph $\mathbb E[\overline Q]$, 
from the result of Lemma~\ref{lem:en-4nodes-delta-quasi}, we have that
$$
\mathbb E[Q]=\frac{\frac{(0.5-2\delta)(0.5+2\delta)}{(4\delta)^2}+\frac{(0.5-\delta)(0.5+\delta)}{(2\delta)^2}+\frac{3\delta(1-3\delta)}{(1-6\delta)^2}}
{1+\frac{0.5-2\delta}{4\delta}+\frac{0.5-\delta}{2\delta}+\frac{3\delta}{1-6\delta}}.
$$
and from the result Lemma~\ref{lem:en-4nodes-delta-complete}
$$
\mathbb E[\overline Q]=\frac{2\frac{(0.25-\delta)(0.75+\delta)}{(0.5+2\delta)^2}+\frac{(0.5-\delta)(0.5+\delta)}{(2\delta)^2}+\frac{3\delta(1-3\delta)}{(1-6\delta)^2}}
{1+2\frac{0.25-\delta}{0.5+2\delta}+\frac{0.5-\delta}{2\delta}+\frac{3\delta}{1-6\delta}}
$$

The desired result follows if we show that
\begin{itemize}
\item[(a)] $\delta^2\left(\frac{(0.5-2\delta)(0.5+2\delta)}{(4\delta)^2}+\frac{(0.5-\delta)(0.5+\delta)}{(2\delta)^2}+\frac{3\delta(1-3\delta)}{(1-6\delta)^2}\right)$ tends to 
$5\cdot 0.5^6$ when $\delta\to 0$,
\item[(b)] $\delta^2\left(2\frac{(0.25-\delta)(0.75+\delta)}{(0.5+2\delta)^2}+\frac{(0.5-\delta)(0.5+\delta)}{(2\delta)^2}+\frac{3\delta(1-3\delta)}{(1-6\delta)^2}\right)$ tends to 
$0.5^4$ when $\delta\to 0$,
\item[(c)] $\tfrac{1}{\delta}\left(1+\frac{0.5-2\delta}{4\delta}+\frac{0.5-\delta}{2\delta}+\frac{3\delta}{1-6\delta}\right)^{-1}$ 
tends to $\tfrac{8}{3}$ when $\delta\to 0$,
\item[(d)] $\tfrac{1}{\delta}\left(1+2\frac{0.25-\delta}{0.5+2\delta}+\frac{0.5-\delta}{2\delta}+\frac{3\delta}{1-6\delta}\right)^{-1}$ tends 
to $4$ when $\delta\to 0$,
\end{itemize}
since when $\delta\to0$
$$
\delta\left(\mathbb E[\overline Q]-\mathbb E[Q]\right)\to0.5^4\cdot 4 -5\cdot 0.5^6\cdot \frac{8}{3}= 0.041\wideparen{6}.
$$
We first show (a). 
\begin{multline*}
\delta^2\left(\frac{(0.5-2\delta)(0.5+2\delta)}{(4\delta)^2}+
\frac{(0.5-\delta)(0.5+\delta)}{(2\delta)^2}+
\frac{3\delta(1-3\delta)}{(1-6\delta)^2}\right)\\
=\frac{(0.5-\delta)(0.5+2\delta)}{16}+
\frac{(0.5-\delta)(0.5+\delta)}{4}+
\frac{3\delta^3(1-3\delta)}{(1-6\delta)^2},
\end{multline*}
and the first and second terms tend, respectively, to $0.5^6$ and $0.5^4 = 4\cdot 0.5^6$ when $\delta\to0$, whereas the third one to zero.
\newline
We now show (b). 
\begin{multline*}
\delta^2\left(2\frac{(0.25-\delta)(0.75+\delta)}{(0.5+2\delta)^2}+
\frac{(0.5-\delta)(0.5+\delta)}{(2\delta)^2}+
\frac{3\delta(1-3\delta)}{(1-6\delta)^2}\right)\\
=2\delta^2\frac{(0.25-\delta)(0.75+\delta)}{(0.5+2\delta)^2}+
\frac{(0.5-\delta)(0.5+\delta)}{4}+
\frac{3\delta^3(1-3\delta)}{(1-6\delta)^2},
\end{multline*}
and the first and third terms tend to zero when $\delta\to0$, whereas the second one to $0.5^4$.
\newline
We also show (c).
\begin{multline*}
\frac{1}{\delta}\left(1+\frac{0.5-2\delta}{4\delta}+
\frac{(0.5-\delta)}{2\delta}+
\frac{3\delta}{1-6\delta}\right)^{-1}\\
=\left(\delta+\frac{(0.5-2\delta)}{4}+
\frac{(0.5-\delta)}{2}+
\frac{3\delta^2}{1-6\delta}\right)^{-1}, \end{multline*}
and when $\delta\to0$, the last expression tends to $\left(\frac{1}{8}+\frac{1}{4}\right)^{-1}$, 
which equals $\frac{8}{3}$.
\newline
Finally, we show (d).
\begin{multline*}
\frac{1}{\delta}\left(1+2\frac{(0.25-\delta)}{0.5+2\delta}+
\frac{(0.5-\delta)}{2\delta}+
\frac{3\delta}{1-6\delta}\right)^{-1}\\
=\left(\delta+2\delta\frac{(0.25-\delta)}{(0.5+2\delta)}+
\frac{(0.5-\delta)}{2}+
\frac{3\delta^2}{1-6\delta}\right)^{-1},
 \end{multline*}
where all the terms tend to zero when $\delta\to0$, except for $\frac{0.5-\delta}{2}$, which 
tends to $0.25$ and, therefore, the desired result follows.

\section{Proof of Proposition~\ref{prop:finite_sums}}
\label{app:prop:finite_sums}

In \cite{MBM20}, the authors prove that $\pi_0=\left(1+\sum_{\I\in\Iset}T_{\I}\right)^{-1}$. Thus, we only need to prove that $\sum_{w\in\mathbb{W}}|w|\cdot\pi(w)=\sum_{\I\in\Iset}E_{\I}$.
First, we rewrite the infinite sum over all possible words into a finite sum over all independent sets:
\begin{equation}\label{eq:pi_bar_0}
\E[Q] = \pi_0 \sum_{w\in\mathbb{W}}|w|\cdot\pi(w)= \pi_0 \sum_{\I\in\Iset}\overline{\pi}_{\I}
\end{equation}
where $\overline{\pi}_{\I}$ is the sum of $|w|\cdot\pi(w)$ over all the words $w$ such that the alphabet of $w$ is exactly $\I$, i.e the set of its unique letters is exactly $\I$. 

Then, we are going to prove that $\overline{\pi}_{\I}=E_{\I}$. Let $\I^o=\{i_1,\cdots,i_{\I}\}$ be an ordered version of $\I$ and $\I^o_k=\{i_1,\cdots,i_k\}$ be the first $k$ elements of $\I^o$ for any $k\in\{1,\cdots,|\I|\}$. We first consider $\overline{\pi}_{\I^o}$ which we define as $\overline{\pi}_{\I}$ where we restricted the sum to be only over the words such that the first appearance of their letters is ordered as in $\I^o$. Let $w$ be such a word, we know that the first letter can only be $i_1$, then we can have any number of times (which we will note $p_1^1$) the letter $i_1$ before the first appearance of the letter $i_2$. After that, there can be any number of times the letters $i_1$ and $i_2$ (which we will note $p_1^2$ and $p_2^2$) before the first appearance of the letter $i_3$. We can repeat this reasoning until the first appearance of the letter $i_{|\I|}$ and then there can be any number of times the letters $i_1,\cdots, i_{|\I|}$ (which we will note $p_1^{|\I|},\cdots, p_{|\I|}^{|\I|}$). At each step $k\in\{1,\cdots,|\I|\}$, we denote by $n_k$ the number of letters between the first appearance of $i_k$ and the first appearance of $i_{k+1}$ (both excluded), i.e $n_k=p_1^k +\cdots +p_k^k$. For $p_1^k,\cdots,p_k^k$ fixed, the stationary distribution is the same regardless of the order of the letters. Thus, we only need to count the number of words of size $n_k$ on an alphabet of size $k$ with each letter $i_l$ repeated $p_l^k$ times which is equal to the multinomial coefficient:
\[\binom{n_k}{p_1^k,\cdots,p_k^k}=\frac{n_k !}{\prod_{l=1}^k p_l^k !}\]
In the end, to get all the words $w$ belonging to the sum in $\overline{\pi}_{\I^o}$, we only need to sum over all possible values of $n_k$ and $p_l^k$ for all $l\in\{1,\cdots,k\}$ and for all $k\in\{1,\cdots,|\I|\}$. Using the definition of $\pi(w)$ as in Proposition~\ref{prop:stationary_dist}, we can compute the value of $\overline{\pi}_{\I^o}$ which is equal to
\begin{align*}
\overline{\pi}_{\I^o}&=\prod_{k=1}^{|\I|}\left(\frac{\alpha_{i_k}}{|\alpha_{\Ecal(\I^o_k)}|}\right) \sum_{n_k=0}^\infty \sum_{p_1^k +\cdots +p_k^k =0}^{n_k}\prod_{m=1}^{k}\left(\frac{\alpha_{i_m}}{|\alpha_{\Ecal(\I^o_k)}|}\right)^{p_m^k}\binom{n_k}{p_1^k,\cdots,p_k^k} \\
&\quad \times\left(|\I|+\sum_{k=1}^{|\I|}n_k \right) \\
&=\prod_{k=1}^{|\I|}\left(\frac{\alpha_{i_k}}{|\alpha_{\Ecal(\I^o_k)}|}\right) \sum_{n_k=0}^\infty \left(\frac{|\alpha_{\I^o_k}|}{|\alpha_{\Ecal(\I^o_k)}|}\right)^{n_k}  \left(|\I|+\sum_{k=1}^{|\I|}n_k \right) \\
&=\sum_{l=1}^{|\I|}\prod_{k=1}^{|\I|}\left(\frac{\alpha_{i_k}}{|\alpha_{\Ecal(\I^o_k)}|}\right) \sum_{n_k=0}^\infty \left(\frac{|\alpha_{\I^o_k}|}{|\alpha_{\Ecal(\I^o_k)}|}\right)^{n_k}  (n_l + 1) \\
&=\sum_{l=1}^{|\I|}\left(\frac{\alpha_{i_l}}{|\alpha_{\Ecal(\I^o_l)}|}\right) \sum_{n_l=0}^\infty \left(\frac{|\alpha_{\I^o_l}|}{|\alpha_{\Ecal(\I^o_l)}|}\right)^{n_l}  (n_l + 1) \\
&\quad\times \prod_{\substack{k\in\llbracket 1,|\I|\rrbracket,\\ k\neq l}}\left(\frac{\alpha_{i_k}}{|\alpha_{\Ecal(\I^o_k)}|}\right) \sum_{n_k=0}^\infty \left(\frac{|\alpha_{\I^o_k}|}{|\alpha_{\Ecal(\I^o_k)}|}\right)^{n_k} \\
&=\sum_{l=1}^{|\I|}\left(\frac{\alpha_{i_l}}{|\alpha_{\Ecal(\I^o_l)}|}\right) \left(\frac{|\alpha_{\Ecal(\I^o_l)}|}{|\alpha_{\Ecal(\I^o_l)}|-|\alpha_{\I^o_l}|}\right)^2 \\
&\quad\times \prod_{\substack{k\in\llbracket 1,|\I|\rrbracket,\\ k\neq l}}\left(\frac{\alpha_{i_k}}{|\alpha_{\Ecal(\I^o_k)}|}\right) \left(\frac{|\alpha_{\Ecal(\I^o_k)}|}{|\alpha_{\Ecal(\I^o_k)}|-|\alpha_{\I^o_k}|}\right)  \\
&=\sum_{l=1}^{|\I|}\frac{|\alpha_{\Ecal(\I^o_l)}|}{|\alpha_{\Ecal(\I^o_l)}|-|\alpha_{\I^o_l}|} \prod_{k=1}^{|\I|} \frac{\alpha_{i_k}}{|\alpha_{\Ecal(\I^o_k)}|-|\alpha_{\I^o_k}|} \\
&=E_{\I^o}
\end{align*}

Then, we sum on all permutation of the initial order to get the desired result: $\overline{\pi}_{\I}=\sum_{\sigma\in\mathfrak{S}_{|\I|}}\overline{\pi}_{\I^{\sigma(o)}}=\sum_{\sigma\in\mathfrak{S}_{|\I|}}E_{\I^{\sigma(o)}}=E_{\I}$.

The proof for $\pi_0$ being equal to $\left(1+\sum_{\I\in\Iset}T_{\I}\right)^{-1}$ can be found in \cite{MBM20} but can also be retrieved here by removing the term $\left(|\I| + \sum_{k=1}^{|\I|}n_k\right)$ in the first equation above and then moving on with the computations in a similar way.

\section{Proof of Lemma~\ref{lem:pos_cst_in_sums}}
\label{app:lem:pos_cst_in_sums}

We are first going to look at the proof for $A_2$.
Let $T=1+\sum_{\I\in \Iset} T_{\I}$. We rewrite $T$ such as all terms of the sum are on the same denominator, this can be done naively by multiplying each term by the denominator of all others which would be equal to 
\begin{align*}
\prod_{\I\in \Iset}\prod_{\sigma\in\mathfrak{S}_{|\I|}}\prod_{k=1}^{|\I|}\left(|\alpha_{\Ecal(\I^{\sigma(o)}_k)}|-|\alpha_{\I^{\sigma(o)}_k}|\right)
\end{align*}
where $\I^o=\{i_1,\cdots,i_{|\I|}\}$ is an ordered version of $\I$ and $\I_k^{\sigma(o)}=\{i_{\sigma(1)},\cdots,i_{\sigma(k)}\}$.
However, there is a lot of duplicates in this denominator which could have been avoided. First, for all independent set $\I\in\Iset$, all the terms $\left(|\alpha_{\Ecal(\I^{\sigma(o)}_k)}|-|\alpha_{\I^{\sigma(o)}_k}|\right)$ for $k\in\{1,\cdots,|\I|-1\}$ appear as the last term $\left(|\alpha_{\Ecal(\tilde{\I}^{\sigma(o)})}|-|\alpha_{\tilde{\I}^{\sigma(o)}}|\right)$ of the independent set $\tilde{\I}=\I_k$.
Then, for all independent set $\I\in\Iset$, the last term $\left(|\alpha_{\Ecal(\I^{\sigma(o)})}|-|\alpha_{\I^{\sigma(o)}}|\right)$ will have the same value no matter what is the permutation $\sigma\in\mathfrak{S}_{|\I|}$. Removing the duplicates from both observations lead to the following common denominator for $T$:
\begin{align}\label{eq:T_denominator}
\prod_{\I\in \Iset}\left(|\alpha_{\Ecal(\I)}|-|\alpha_{\I}|\right).
\end{align} 
Thus, $T$ is a rational fraction of $\delta$ as we are only adding and multiplying polynomials in $\delta$.
For the term equal to $1$, the numerator will be equal to the denominator. This means that the constant term of the polynomial in $\delta$ at the numerator is equal to $\prod_{\I\in \Iset}\left(|\alpha_{\Ecal(\I)}|-|\alpha_{\I}|\right)_a$ which is equal to zero because of Assumption~\ref{ass:saturated}.

Let $\tilde{\I}\in \Iset$, we first want to put every term in $T_{\tilde{\I}}$ on the same denominator. For each $T_{\tilde{\I}^{\sigma(o)}}$, we know that there is already all the terms related to $\Iset_{\tilde{\I}^{\sigma(o)}}=\left\{\tilde{\I}^{\sigma(o)}_k, \forall k\in\{1,\cdots,|\tilde{\I}|\}\right\}$ at the denominator. The terms missing are all the others which are related to $\Iset_{\tilde{\I}}$. Then, for each $T_{\tilde{\I}}$ the terms missing at the denominator are all the others related to $\Iset$. This means that the constant term of the polynomial in $\delta$ at the numerator of $T_{\tilde{\I}}$ is equal to
\begin{align}\label{eq:T_constant_num}
\prod_{i\in\tilde{\I}}a_{i}\left(\sum_{\sigma\in\mathfrak{S}_{|\tilde{\I}|}}\prod_{\I\in \Iset_{\tilde{\I}}\setminus \Iset_{\tilde{\I}^{\sigma(o)}}}\left(|\alpha_{\Ecal(\I)}|-|\alpha_{\I}|\right)_a\right) \prod_{\I\in \Iset\setminus\Iset_{\tilde{\I}}}\left(|\alpha_{\Ecal(\I)}|-|\alpha_{\I}|\right)_a
\end{align} 
and is non-negative because of Assumption~\ref{ass:saturated}. 
Let $\tilde{\I}=\hat{\I}$, it is obvious that $\hat{\I}\in\Iset_{\hat{\I}}$ which means that the rightmost term in \eqref{eq:T_constant_num} is positive (because $\left(|\alpha_{\Ecal(\I)}|-|\alpha_{\I}|\right)_a$ is equal to zero only for $\I=\hat{\I}$ by Assumption~\ref{ass:saturated}). In addition, 
for all the permutation $\sigma$ we have $\hat{\I}^{\sigma(o)}_{|\hat{\I}|}=\hat{\I}$, i.e $\hat{\I}\in\Iset_{\hat{\I}^{\sigma(o)}}$ which means that the middle term is also positive. Thus, the constant term described in \eqref{eq:T_constant_num} is positive for $\tilde{\I}=\hat{\I}$ (but can also be positive for other $\tilde{\I}$). To conclude, because we are only adding non-negative constant terms and because there exists at least one positive constant term, the polynomial in $\delta$ at the numerator of $T$ has a positive constant term. All the arguments remain valid if we replace $T_{\I}$ by $\overline{T}_{\I}$ and $\Iset$ by $\overline{\Iset}$ giving the desired result for $B_2$.

A similar proof can be made for $E=\sum_{\I\in I}E_{\I}$ for any subset $I\in\Iset$ with the following common denominator for $E$:
\begin{align}\label{eq:E_denominator}
\prod_{\I\in I^+}\left(|\alpha_{\Ecal(\I)}|-|\alpha_{\I}|\right)^2 
\end{align}
where $I_+=\bigcup_{\I\in I}\Iset_{\I}$.
Let $\tilde{\I}\in I$, the constant term of the polynomial in $\delta$ at the numerator of $E_{\tilde{\I}}$ is equal to
\begin{align}\label{eq:E_constant_num}
&\prod_{i\in\tilde{\I}}a_{i}\left(\sum_{\sigma\in\mathfrak{S}_{|\tilde{\I}|}}\left(\sum_{\dot{\I}\in\Iset_{\tilde{\I}^{\sigma(o)}}}\left(|\alpha_{\Ecal(\dot{\I})}|\right)_a \prod_{\I\in\Iset_{\tilde{\I}^{\sigma(o)}}\setminus\{\dot{\I}\}}\left(|\alpha_{\Ecal(\I)}|-|\alpha_{\I}|\right)_a\right) \right.\nonumber\\
&\left.\times \prod_{\I\in \Iset_{\tilde{\I}}\setminus \Iset_{\tilde{\I}^{\sigma(o)}}}\left(|\alpha_{\Ecal(\I)}|-|\alpha_{\I}|\right)_a^2\right) \prod_{\I\in I_+\setminus\Iset_{\tilde{\I}}}\left(|\alpha_{\Ecal(\I)}|-|\alpha_{\I}|\right)_a^2
\end{align} 
Choose $I=\Iset^\ast$ to get the proof for $B_1$ and $I=\Iset^{-\ast}$ to get the proof for $C_1$. All the arguments remain valid if we replace $E_{\I}$ by $\overline{E}_{\I}$ and $\Iset$ by $\overline{\Iset}$ giving the desired result for $A_1$ when $I=\overline{\Iset}^\ast$.

\section{Proof of Lemma~\ref{lem:pos_cst_in_sums_C}}
\label{app:lem:pos_cst_in_sums_C}

Let $\Iset^\ast_+=\bigcup_{\I\in\Iset^\ast}\Iset_{\I}$. We recall that $\overline{\Iset}^\ast\subseteq \Iset^\ast$ because adding an edge can only decrease the size of independent sets.
Let $\tilde{\I}\in\overline{\Iset}^\ast$, we are interested in the term $T_{\tilde{\I}}-\overline{T}_{\tilde{\I}}$. We start by putting that term on the same denominator which is equal to
\[
\prod_{\I\in\overline{\Iset}_{\tilde{\I}}\cap\Iset^{-\ast}}\left(|\alpha_{\Ecal(\I)}|-|\alpha_{\I}|\right)\prod_{\I\in\overline{\Iset}_{\tilde{\I}}\cap\overline{\Iset}^\ast}\left(|\alpha_{\Ecal(\I)}|-|\alpha_{\I}|\right)\left(|\alpha_{\overline{\Ecal}(\I)}|-|\alpha_{\I}|\right)
\]
In order to compute the constant term of the polynomial in $\delta$ at the numerator of $T_{\tilde{\I}}-\overline{T}_{\tilde{\I}}$, we will use a similar proof as in Lemma~\ref{lem:pos_cst_in_sums}. In particular, we are going to use \eqref{eq:T_constant_num} without the rightmost term (because we only want a common denominator for $T_{\tilde{\I}}-\overline{T}_{\tilde{\I}}$ and not the whole $C_2$). First, we separate the product on $\I\in \Iset_{\tilde{\I}}\setminus \Iset_{\tilde{\I}^{\sigma(o)}}$ in two because if $\I\in\overline{\Iset}^\ast$ then we have to distinguish the term related to $\Ecal(\I)$ and the one related to $\overline{\Ecal}(\I)$, otherwise $\I\in\Iset^{-\ast}$ and $\Ecal(\I)=\overline{\Ecal}(\I)$. In addition, to have $T_{\tilde{\I}}-\overline{T}_{\tilde{\I}}$ on the same denominator, we have to multiply the numerator (and the denominator) of $T_{\tilde{\I}}$ by $\prod_{\I\in\overline{\Iset}_{\tilde{\I}}\cap\overline{\Iset}^{\ast}}\left(|\alpha_{\overline{\Ecal}(\I)}|-|\alpha_{\I}|\right)$ and of $\overline{T}_{\tilde{\I}}$ by $\prod_{\I\in\overline{\Iset}_{\tilde{\I}}\cap\overline{\Iset}^{\ast}}\left(|\alpha_{\Ecal(\I)}|-|\alpha_{\I}|\right)$. Finally, if $\I\in \left(\overline{\Iset}_{\tilde{\I}}\cap \overline{\Iset}^{\ast}\right)\setminus \overline{\Iset}_{\tilde{\I}^{\sigma(o)}}$, then the numerator of $T_{\tilde{\I}}$ and the numerator of $\overline{T}_{\tilde{\I}}$ will both be multiplied by the terms related to $\Ecal(\I)$ and $\overline{\Ecal}(\I)$ which means that we can factorize them and the difference in $T_{\tilde{\I}}-\overline{T}_{\tilde{\I}}$ will only appear for $\I\in \overline{\Iset}_{\tilde{\I}^{\sigma(o)}}\cap\overline{\Iset}^{\ast}$. All these observations lead to the constant term being equal to
\begin{align}\label{eq:T_constant_num_C}
&\prod_{i\in\tilde{\I}}a_{i}\left(\sum_{\sigma\in\mathfrak{S}_{|\tilde{\I}|}}\prod_{\I\in \left(\overline{\Iset}_{\tilde{\I}}\cap \Iset^{-\ast}\right)\setminus \overline{\Iset}_{\tilde{\I}^{\sigma(o)}}}\left(|\alpha_{\Ecal(\I)}|-|\alpha_{\I}|\right)_a \right.\nonumber\\
&\times \prod_{\I\in \left(\overline{\Iset}_{\tilde{\I}}\cap \overline{\Iset}^{\ast}\right)\setminus \overline{\Iset}_{\tilde{\I}^{\sigma(o)}}}\left(|\alpha_{\Ecal(\I)}|-|\alpha_{\I}|\right)_a \left(|\alpha_{\overline{\Ecal}(\I)}|-|\alpha_{\I}|\right)_a \nonumber\\
&\left.\times\prod_{\I\in  \overline{\Iset}_{\tilde{\I}^{\sigma(o)}}\cap\overline{\Iset}^{\ast}}\left[\left(|\alpha_{\overline{\Ecal}(\I)}|-|\alpha_{\I}|\right)_a - \left(|\alpha_{\Ecal(\I)}|-|\alpha_{\I}|\right)_a\right]\right) 
\end{align} 
which is non-negative because of Assumption~\ref{ass:saturated} and because the rightmost term is non-negative as $\Ecal(\I)\subseteq \overline{\Ecal}(\I)$ for all $\I\in\overline{\Iset}^{\ast}$.

Let $T=\sum_{\I\in\Iset^{\ast}\setminus\overline{\Iset}^{\ast}}T_{\I}$, we can use a similar proof as in Lemma~\ref{lem:pos_cst_in_sums} to show that $T$ can be written as a rational fraction of $\delta$ such that the polynomial at the numerator has a positive constant term if $\hat{\I}\notin \Iset^\ast_+$. 

Using a similar proof as in Lemma~\ref{lem:pos_cst_in_sums}, we can rewrite $C_2$ such that all terms are on the same denominator which is equal to
\[
\prod_{\I\in\Iset^\ast_+\setminus\Iset^\ast}\left(|\alpha_{\Ecal(\I)}|-|\alpha_{\I}|\right)\prod_{\I\in\Iset^\ast}\left(|\alpha_{\Ecal(\I)}|-|\alpha_{\I}|\right)\prod_{\I\in\overline{\Iset}^\ast}\left(|\alpha_{\overline{\Ecal}(\I)}|-|\alpha_{\I}|\right)
\]
where the leftmost term correspond to the independent sets $\I$ that do not contain $i^\ast$ or $j^\ast$, thus having $\Ecal(\I)=\overline{\Ecal}(\I)$. Therefore, the constant terms at the numerator of $T$ and of $T_{\tilde{\I}}-\overline{T}_{\tilde{\I}}$ are multiplied by positive terms if $\hat{\I}\notin \Iset^\ast_+$. To conclude, because we are only adding non-negative constant terms and because there exists at least one positive constant term, the polynomial in $\delta$ at the numerator of $C_2$ has a positive constant term.

\end{document}